\documentclass[twocolumn,showpacs,preprintnumbers,amsmath,amssymb]{revtex4}

\usepackage{graphicx}
\usepackage{dcolumn}
\usepackage{bm}

\usepackage{amssymb}
\usepackage{amsmath}

\begin{document}


\title{Entropic cosmology for a generalized black-hole entropy}

\author{Nobuyoshi {\sc Komatsu}$^{1}$}  \altaffiliation{E-mail: komatsu@t.kanazawa-u.ac.jp} 
\author{Shigeo     {\sc Kimura}$^{2}$}

\affiliation{$^{1}$Department of Mechanical Systems Engineering, Kanazawa University, 
                          Kakuma-machi, Kanazawa, Ishikawa 920-1192, Japan \\
                $^{2}$The Institute of Nature and Environmental Technology, Kanazawa University, 
                          Kakuma-machi, Kanazawa, Ishikawa 920-1192, Japan}%
\date{\today}

\begin{abstract}
An entropic-force scenario, i.e., entropic cosmology, assumes that the horizon of the universe has an entropy and a temperature.
In the present study, in order to examine entropic cosmology, we derive entropic-force terms not only from the Bekenstein entropy 
but also from a generalized black-hole entropy proposed by C. Tsallis and L.J.L. Cirto [Eur. Phys. J. C \textbf{73}, 2487 (2013)]. 
Unlike the Bekenstein entropy, which is proportional to area, the generalized entropy is proportional to volume because of appropriate nonadditive generalizations. 
The entropic-force term derived from the generalized entropy is found to behave as if it were an extra driving term for bulk viscous cosmology, in which a bulk viscosity of cosmological fluids is assumed. 
Using an effective description similar to bulk viscous cosmology, we formulate the modified Friedmann, acceleration, and continuity equations for entropic cosmology. 
Based on this formulation, we propose two entropic-force models derived from the Bekenstein and generalized entropies. 
In order to examine the properties of the two models, we consider a homogeneous, isotropic, and spatially flat universe, focusing on a single-fluid-dominated universe. 
The two entropic-force models agree well with the observed supernova data. 
Interestingly, the entropic-force model derived from the generalized entropy predicts a decelerating and accelerating universe, as for a fine-tuned standard $\Lambda$CDM (lambda cold dark matter) model, 
whereas the entropic-force model derived from the Bekenstein entropy predicts a uniformly accelerating universe.
\end{abstract}

\pacs{98.80.-k, 98.80.Es, 95.30.Tg}
\maketitle

\section{Introduction}

Since the late 1990s, an accelerated expansion of the late universe has gradually been accepted as a new paradigm \cite{PERL1998ab,Riess1998_2004,Riess2007SN1,Tegmark1_many,WMAP2011,Planck2013}. 
In order to explain this accelerated expansion, various cosmological models have been suggested \cite{Ryden1,Weinberg1,Roy1,Miao1,Bamba1}.
(See, e.g., Refs.\ \cite{Weinberg1,Roy1,Miao1,Bamba1} and the references therein.)
As one such model, Easson, Frampton, and Smoot \cite{Easson1,Easson2} recently proposed an entropic-force scenario called `entropic cosmology'.
In entropic cosmology, an extra driving term, i.e., an entropic-force term, should be added to the Friedmann--Lema\^{i}tre acceleration equation, without introducing new fields \cite{Koivisto1}.
The entropic-force term is derived from the usually neglected surface terms on the horizon of the universe, assuming that the horizon has an entropy and a temperature due to the information holographically stored there \cite{Easson1}.  

Entropic cosmology has been extensively examined from various viewpoints \cite{Koma4,Koma4bc,Cai1,Cai2,Qiu1,Casadio1-Costa1,Basilakos1}. 
As an entropy on the horizon, i.e., the Bekenstein entropy \cite{Bekenstein1} is always used, substituting the horizon of the universe for the event horizon of a black hole.   
In fact, the Bekenstein entropy is proportional to area (or horizon) and is additive.
However, self-gravitating systems exhibit peculiar features \cite{BT1,Tsa1}, 
such as nonequilibrium thermodynamics and nonextensive statistical mechanics  \cite{Lebo1-Thir4,Ren1,Tsa0,Plas1,Abe1,Torres1,Tsa101,Tsa2,Chava21,Taru4,Naka1,Taru5,Liu1,Koma2,Koma3,Everton1,Tsallis2012}. 
Accordingly, for example, the Tsallis entropy \cite{Tsa0} and the Renyi entropy \cite{Ren1} have been proposed for nonadditive (nonextensive) generalized entropies and have been investigated from astrophysical viewpoints \cite{Plas1,Abe1,Torres1,Tsa101,Tsa2,Chava21,Taru4,Naka1,Taru5,Liu1,Koma2,Koma3,Everton1,Tsallis2012}.   
In particular, Tsallis and Cirto recently suggested a generalized black-hole entropy proportional to its volume based on appropriate nonadditive generalizations \cite{Tsallis2012}.  
Using the generalized entropy instead of the Bekenstein entropy will provide new insight into entropic cosmology. 
Therefore, it is important to derive entropic-force models from the two entropies and examine the properties of the two models in order to understand entropic cosmology more deeply.  
Note that power-law and logarithmic entropic-corrections have been discussed  \cite{Easson2,Koivisto1,Cai1,Cai2,Qiu1,Sheykhi1,Sadjadi1}.

In addition, in entropic cosmology, the entropy on the horizon of the universe can increase during the evolution of the universe \cite{Koma4}, even if we consider a homogeneous and isotropic universe.  
However, a bulk viscosity of cosmological fluids \cite{Davies3,Weinberg0,Murphy1,Barrow11,Barrow12,Lima101,Zimdahl1,Arbab1,Brevik1,Brevik2,Nojiri1,Meng1,Fabris1,Colistete1,Barrow21,Meng2,Avelino1,Hipo1,Avelino2,Piattella1,Meng3,Pujolas1,Odintsov1,Odintsov2,Odintsov3,Odintsov4} is usually the only thing that can generate an entropy in the homogeneous and isotropic universe \cite{Lima101}.  
(Such a cosmological model is referred to as bulk viscous cosmology.) 
Through the study of entropic cosmology, we may be able to discuss the classical entropy generated by bulk viscous stresses.

In this context, we examine entropic cosmology using a generalized black-hole entropy (proportional to its volume) proposed by Tsallis and Cirto \cite{Tsallis2012}. 
In the present study, we derive entropic-force terms not only from the Bekenstein entropy but also from the generalized entropy.  
Moreover, using an effective description similar to bulk viscous cosmology, we formulate the modified Friedmann, acceleration, and continuity (conservation) equations for entropic cosmology. 
Note that the entropic-force considered here is different from the idea that gravity itself is an entropic force \cite{Padma1,Verlinde1}.  
Since we focus on background evolutions of the late universe, we do not discuss the inflation of the early universe.

The remainder of the present paper is organized as follows.
In Sec.\ \ref{Cosmological models}, we present a brief review of three cosmological models, i.e., 
$\Lambda$CDM (lambda cold dark matter) cosmology, bulk viscous cosmology, and entropic cosmology. 
In Sec.\ \ref{Derivation of entropic-force}, we derive entropic-force terms from both the Bekenstein entropy and a generalized black-hole entropy.  
We also formulate the modified Friedmann, acceleration, and continuity equations for two entropic-force models based on the obtained entropic-force terms.
In Sec.\ \ref{combined model}, we examine a model that combines the two entropic-force models. 
In Sec.\ \ref{Comparison}, we discuss the evolution of the universe in the two entropic-force models using solutions of the combined model.
Finally, in Sec.\ \ref{Conclusions}, we present a discussion and our conclusions.

\section{Cosmological models}
\label{Cosmological models}
In the present paper, we consider a homogeneous, isotropic, and spatially flat universe 
and examine the scale factor $a(t)$ at time $t$ in the Friedmann--Lema\^{i}tre--Robertson--Walker metric.
In this section, we present a brief review of three cosmological models, i.e., $\Lambda$CDM, bulk viscous, and entropic cosmologies, 
focusing on the Friedmann, acceleration, and continuity equations.
(A spatially non-flat universe is discussed in the last paragraph of in Sec.\ \ref{combined model}.)

\subsection{$\Lambda$CDM cosmology} 
\label{LCDM models}
We first introduce the well-known $\Lambda$CDM models \cite{Weinberg1,Ryden1,Roy1}.
In the standard $\Lambda$CDM model, the Friedmann equation is given as 
\begin{equation}
  \left(  \frac{ \dot{a}(t) }{ a(t) } \right)^2  =  H(t)^2     =  \frac{ 8\pi G }{ 3 } \rho(t) + \frac{\Lambda}{3} ,  
\label{eq:FRW1_LCDM}
\end{equation}
and the acceleration equation is 
\begin{equation}
  \frac{ \ddot{a}(t) }{ a(t) }   =  \dot{H}(t) + H(t)^{2}   
                                          =  -  \frac{ 4\pi G }{ 3 }  \left (  \rho(t) +  \frac{3p(t)}{c^2}  \right )  + \frac{\Lambda}{3}.
\label{eq:FRW2_LCDM}
\end{equation}
In addition, the continuity equation is given by 
\begin{equation}
 \dot{\rho}(t) + 3 \frac{\dot{a}(t)}{a(t)} \left (  \rho (t) + \frac{p(t)}{c^2}  \right ) = 0  ,
\label{eq:fluid_LCDM}
\end{equation}
where the Hubble parameter $H(t)$ is defined by
\begin{equation}
   H(t) \equiv   \frac{ da/dt }{a(t)} =   \frac{ \dot{a}(t) } {a(t)}  ,
\label{eq:Hubble}
\end{equation}
and $G$, $\Lambda$, $c$, $\rho(t)$, and $p(t)$ are the gravitational constant, a cosmological constant, the speed of light, the mass density of cosmological fluids, and the pressure of cosmological fluids, respectively. Equations (\ref{eq:FRW1_LCDM}) and  (\ref{eq:FRW2_LCDM}) include the extra driving terms, i.e., $\Lambda/3$, which can explain the accelerated expansion of the late universe.

The continuity equation is consistent with the Friedmann and acceleration equations, because two of the three equations are independent \cite{Ryden1}.
In other words, if the Friedmann and acceleration equations are used, Eq.\ (\ref{eq:fluid_LCDM}) is derived from these two equations.
Note that Eq.\ (\ref{eq:fluid_LCDM}) can be derived from the first law of thermodynamics as well, assuming adiabatic (isentropic) processes, without using the Friedmann and acceleration equations \cite{Ryden1}.

\subsection{Bulk viscous cosmology} 
\label{Bulk viscous models}
In bulk viscous cosmology, a bulk viscosity $\eta$ of cosmological fluids is assumed 
\cite{Davies3,Weinberg0,Murphy1,Barrow11,Barrow12,Lima101,Zimdahl1,Arbab1,Brevik1,Brevik2,Nojiri1,Meng1,Fabris1,Colistete1,Barrow21,Meng2,Avelino1,Hipo1,Avelino2,Piattella1,Meng3,Pujolas1,Odintsov1,Odintsov2,Odintsov3,Odintsov4}. Usually, the bulk viscosity is the only thing that can generate an entropy in the homogeneous and isotropic universe \cite{Lima101}.
Such a cosmological model is referred to as bulk viscous cosmology. 
For example, in the 1980s, Barrow \cite{Barrow11,Barrow12}, Davies \cite{Davies3}, and Lima \textit{et al}. \cite{Lima101} investigated bulk viscous cosmology in order to discuss the inflation of the early universe.
Recently, a number of studies have examined not only the inflation but also the accelerated expansion of the late universe, based on bulk viscous cosmology 
\cite{Zimdahl1,Arbab1,Brevik1,Brevik2,Nojiri1,Meng1,Fabris1,Colistete1,Barrow21,Meng2,Avelino1,Hipo1,Avelino2,Piattella1,Meng3,Pujolas1,Odintsov1,Odintsov2,Odintsov3,Odintsov4}.

Generally, an effective pressure $p^{\prime}$ for bulk viscous cosmology is given by
\begin{equation}
 p^{\prime} (t) = p(t) - 3 \eta  H(t)    ,  
\label{eq:Bulk_p}
\end{equation} 
where, for simplicity, $\eta$ is a constant.
(It is possible to assume a variable bulk-viscosity. For instance, see Refs.\ \cite{Avelino2,Meng3} and the references therein.)
In bulk viscous cosmology, the Friedmann equation is given by 
\begin{equation}
 \left (  \frac{\dot{a}}{a} \right)^2 = \frac{8\pi G}{3} \rho   .         
\label{eq:Bulk_FRW1}
\end{equation}
Unlike the $\Lambda$CDM model [Eq.\ (\ref{eq:FRW1_LCDM})], Eq. (\ref{eq:Bulk_FRW1}) does not include an extra term such as a cosmological constant.
Using the effective pressure $p^{\prime}$, the acceleration equation for bulk viscous cosmology is given by 
\begin{equation}
\frac{\ddot{a}}{a}  
=  - \frac{4\pi G}{3}  \left (  \rho + 3 \frac{p^{\prime}}{c^2}  \right )  .       
\label{eq:Bulk_FRW2Pprime}
\end{equation}
Substituting Eq.\ (\ref{eq:Bulk_p}) into Eq.\ (\ref{eq:Bulk_FRW2Pprime}) and rearranging, we obtain the acceleration equation as
\begin{equation}
\frac{\ddot{a}}{a}  
=  - \frac{4\pi G}{3}  \left (  \rho + 3 \frac{p}{c^2}  \right ) + \frac{12\pi G }{c^2} \eta H    .
\label{eq:Bulk_FRW2}
\end{equation}
The last term, $12\pi G \eta H / c^2 $, corresponds to the extra driving term \cite{Koma4}.
Accordingly, instead of the cosmological constant, the extra term due to the bulk viscosity can explain the accelerated expansion of the universe.
Note that the extra term considered here is proportional to the Hubble parameter $H(t)$.
The continuity equation for bulk viscous cosmology is given by
\begin{equation}
 \dot{\rho} + 3 \frac{\dot{a}}{a}   \left (  \rho + \frac{p^{\prime}}{c^2}  \right ) = 0 . 
\label{eq:Bulk_fluid0}
\end{equation}
Substituting Eq.\ (\ref{eq:Bulk_p}) into Eq.\ (\ref{eq:Bulk_fluid0}) and rearranging, we have  
\begin{equation}
 \dot{\rho} + 3 \frac{\dot{a}}{a} \left (  \rho + \frac{p}{c^2}  \right ) = \frac{ 9  \eta}{c^2}  H^2  .
\label{eq:Bulk_fluid1}
\end{equation}
Equation (\ref{eq:Bulk_fluid1}) has a non-zero right-hand side related to a classical entropy generated by bulk viscous stresses \cite{Barrow11,Barrow12,Koma4}.

We now derive the continuity equation using a different approach.
To this end, we use the generalized continuity equation obtained from the general Friedmann and acceleration equations \cite{Koma4}, 
because two of the three equations are independent.
The general Friedmann and acceleration equations are given, respectively, by 
\begin{equation}
  \left(  \frac{ \dot{a} }{ a } \right)^2  =  \frac{ 8\pi G }{ 3 } \rho + f(t) ,  
\label{eq:General_FRW01_f}
\end{equation}
and 
\begin{equation}
  \frac{ \ddot{a} }{ a }  =  -  \frac{ 4\pi G }{ 3 } \left ( \rho + \frac{3p}{c^2}  \right ) + g(t) ,
\label{eq:General_FRW02_g}
\end{equation}
where $f(t)$ and $g(t)$ are general functions.
Using Eqs.\ (\ref{eq:General_FRW01_f}) and (\ref{eq:General_FRW02_g}), we obtain the generalized continuity equation \cite{Koma4} given by
\begin{equation}
       \dot{\rho} + 3  \frac{\dot{a}}{a} \left (  \rho + \frac{p}{c^2}  \right )
          =  \frac{3}{4 \pi G} H \left(  - f(t) -  \frac{\dot{f}(t)}{2 H }  +  g(t)      \right )     .
\label{eq:drho_General0}
\end{equation}
Comparing [Eqs.\ (\ref{eq:General_FRW01_f}) and (\ref{eq:General_FRW02_g})] with [Eqs.\ (\ref{eq:Bulk_FRW1}) and (\ref{eq:Bulk_FRW2})], 
we can set the general functions as $f(t) = 0$ and $g(t) = 12\pi G \eta H / c^2 $.
Substituting these functions into Eq.\ (\ref{eq:drho_General0}), we obtain 
\begin{equation}
       \dot{\rho} + 3  \frac{\dot{a}}{a} \left (  \rho + \frac{p}{c^2}  \right )
          =  \frac{3}{4 \pi G} H \left(   \frac{12\pi G}{ c^2} \eta H     \right )   = \frac{ 9  \eta}{c^2}  H^2  .
\label{eq:Bulk_fluid_g1}
\end{equation}
Equation (\ref{eq:Bulk_fluid_g1}) is the same as Eq.\ (\ref{eq:Bulk_fluid1}).
This indicates that Eq.\ (\ref{eq:Bulk_fluid1}) obtained from the effective pressure is consistent with Eq.\ (\ref{eq:Bulk_fluid_g1}) obtained from the Friedmann and acceleration equations.
The consistency plays an important role in formulating the entropic-force models discussed in Sec.\ \ref{Derivation of entropic-force}.

\subsection{Entropic cosmology} 
\label{Standard entropic-force models}
In the entropic cosmology suggested by Easson \textit{et al.} \cite{Easson1,Easson2}, the horizon of the universe is assumed to have an associated entropy and an approximate temperature.
The entropy considered here is the Bekenstein entropy. In the present study, we refer to this cosmology as the standard entropic cosmology. (The Bekenstein entropy is discussed in detail in Sec.\ \ref{Entropic-force from the Bekenstein entropy}.)

In a study by Koivisto {\it et al.} \cite{Koivisto1}, the modified Friedmann and acceleration equations are summarized as 
\begin{equation}
  \left(  \frac{ \dot{a} }{ a } \right)^2     =  \frac{ 8\pi G }{ 3 } \rho  + \alpha_{1} H^2    + \alpha_{2} \dot{H} ,  
\label{eq:SmFRW01}
\end{equation}
and
\begin{equation}
  \frac{ \ddot{a} }{ a }   =  -  \frac{ 4\pi G }{ 3 }  \left (  \rho +  \frac{3p}{c^2}  \right )   + \beta_{1} H^2  + \beta_{2} \dot{H}   .
\label{eq:SmFRW02}
\end{equation}
The four coefficients $\alpha_1$, $\alpha_2$, $\beta_1$, and $\beta_2$ are dimensionless constants.
The $H^2$ and $\dot{H}$ terms with the dimensionless constants correspond to the extra driving terms, i.e., entropic-force terms.
The entropic-force terms are derived, taking into account the entropy and temperature on the horizon of the universe due to the information holographically stored there \cite{Easson1}. 
We can solve the two equations assuming the single-fluid-dominated universe \cite{Koivisto1,Koma4}.
Note that we neglect high-order terms for quantum corrections, because we do not discuss the inflation of the early universe in the present paper.
(The dimensionless constants were expected to be bounded by $3/(2 \pi) \lesssim \beta_{1} \leqslant 1$ and $0 \leqslant \beta_{2} \lesssim 3/(4 \pi) $, 
and typical values for a better fitting were $\beta_{1} = 3/(2 \pi)$ and $\beta_{2} = 3/(4 \pi)$ \cite{Easson1}. 
It was argued that the extrinsic curvature at the surface was likely to result in 
something like $\alpha_1 = \beta_1 = 3/(2 \pi)$ and $\alpha_2 = \beta_2 = 3/(4 \pi)$ \cite{Easson2,Koivisto1}.)

As examined in Ref.\ \cite{Koma4}, we can simplify the two modified Friedmann equations, 
assuming a non-adiabatic-like expansion of the universe. 
The simple modified Friedmann and acceleration equations are summarized as
\begin{equation}
 \left(  \frac{ \dot{a} }{ a } \right)^2    =  \frac{ 8\pi G }{ 3 } \rho  + \alpha_{1} H^2   ,  
\label{eq:SmFRW01s}
\end{equation}
\begin{equation}
\frac{ \ddot{a} }{ a }  =      -  \frac{ 4\pi G }{ 3 } \left ( \rho +  \frac{3p}{c^2} \right )        +  \beta_{1}  H^2  .
\label{eq:SmFRW02s}
\end{equation}
Equations (\ref{eq:SmFRW01s}) and (\ref{eq:SmFRW02s}) do not include $\dot{H}$ terms.
In fact, Easson {\it et al.} first proposed that the entropic-force terms of the modified acceleration equation are $H^2$ or $\frac{3}{2 \pi} H^2$; 
i.e., $\dot{H}$ terms are not included in the entropic-force terms \cite{Easson1}.
In other words, Eq.\ (\ref{eq:SmFRW02s}) is consistent with their original acceleration equation, as discussed in Sec. \ref{Entropic-force from the Bekenstein entropy}.
In the present paper, we select Eqs.\ (\ref{eq:SmFRW01s}) and (\ref{eq:SmFRW02s}) as standard entropic-force models.
(Note that it may be possible to neglect the entropic-force terms of the modified Friedmann equation, i.e., $\alpha_{1}=\alpha_{2} = 0$. 
In the standard entropic cosmology \cite{Koivisto1}, the entropic-force terms are added not only with the acceleration equation but also with the Friedmann equation, as in the case of the $\Lambda$CDM models.)

In Ref.\ \cite{Easson1}, Easson \textit{et al.} considered that the continuity equation was given by $\dot{\rho} + 3 (\dot{a}/{a})  [ \rho + (p/c^{2}) ] = 0$, assuming an adiabatic (isentropic) expansion. 
However, in entropic cosmology, since the entropy on the horizon is assumed, the entropy can increase during the evolution of the universe.
Therefore, in a previous study, we derived the continuity equation from the first law of thermodynamics, 
taking into account a non-adiabatic-like process caused by the entropy and the temperature on the horizon \cite{Koma4}.
Consequently, the modified continuity equation was written as
\begin{equation}
       \dot{\rho} + 3  \frac{\dot{a}}{a} \left ( \rho   +  \frac{p}{c^2} \right )  
          = - \gamma \left( \frac{3}{4 \pi G} H  \dot{H} \right )   .
\label{eq:fluid_thermo(add)}
\end{equation}
Because of the non-zero right-hand side, Eq.\ (\ref{eq:fluid_thermo(add)}) was likely consistent with the continuity equation obtained from the modified Friedmann and acceleration equations \cite{Koma4}.
In the present study, we do not use Eq.\ (\ref{eq:fluid_thermo(add)}) as the modified continuity equation.
Instead, in the next section, we derive the modified continuity equation from an effective pressure for entropic cosmology.

We point out that a similar non-zero right-hand side of the continuity equation appears not only in bulk viscous cosmology but also in `energy exchange cosmology',  
in which the transfer of energy between two fluids is assumed \cite{Koma4,Barrow22},  
e.g., the interaction between matter and radiation \cite{Davidson1Szy1}, matter creation \cite{Lima1112}, 
interacting quintessence \cite{Amendola1Zimdahl01}, the interaction between dark energy and dark matter \cite{Wang0102}, and 
dynamical vacuum energy \cite{Freese1-Fritzsch1,Overduin1,Sola_2002,Sola_2003,Sola_2004,Sola_2009,Sola_2011a,Sola_2011b,Sola_2013a,Sola_2013b,Sola_2013c}.
In particular, cosmological equations for dynamical vacuum energy models are very similar to those for entropic-force models. 
Therefore, in the following paragraph, we discuss a similarity between the two models.

Instead of a cosmological constant, a variable cosmological term $\Lambda(t)$ is assumed in dynamical vacuum energy models, i.e., $\Lambda(t)$CDM models  \cite{Freese1-Fritzsch1,Overduin1,Sola_2002,Sola_2003,Sola_2004,Sola_2009,Sola_2011a,Sola_2011b,Sola_2013a,Sola_2013b,Sola_2013c}. 
Various $\Lambda(t)$ terms (e.g., $H^2$, $H$, and constant terms) have been examined in those works. 
For example, see Refs.\ \cite{ Sola_2002,Sola_2003,Sola_2004,Sola_2009,Sola_2011a}. 
Recently, $H^2$ and $\dot{H}$ terms have been investigated in Refs. \cite{Basilakos1,Sola_2013c}.
[The $H^2$ and $\dot{H}$ terms are the same as the entropic-force terms shown in Eqs. (\ref{eq:SmFRW01}) and (\ref{eq:SmFRW02}).]
Consequently, the influence of $\dot{H}$ terms was found to be similar to that of $H^2$ terms.
In other words, essentially new properties of the late universe were not obtained from $\dot{H}$ terms.
This implies that the $\dot{H}$ terms can be neglected as if Eqs. (\ref{eq:SmFRW01s}) and (\ref{eq:SmFRW02s}) were the approximated Friedmann and acceleration equations, respectively.  
In addition, in $\Lambda(t)$CDM models, the continuity equation for matter `$m$' is arranged as 
$\dot{\rho}_{m}           + 3 (\dot{a} / a) ( \rho_{m}            + p_{m}/c^2 )            = -\dot{\Lambda}$ \cite{Sola_2009}.
The continuity equation is equivalent to Eq.\ (\ref{eq:fluid_thermo(add)}), when $\Lambda(t)$ is proportional to $H^2$.
In this way, cosmological equations for $\Lambda(t)$CDM models are similar to those for entropic-force models.
Accordingly, the cosmological equations and their solutions discussed in the present paper are similar to those for $\Lambda(t)$CDM models, especially in the work of Basilakos, Plionis, and Sol\`{a} \cite{Sola_2009}.
However, a theoretical background of the $\Lambda(t)$CDM model is different from that of the entropic-force model.

Finally, we discuss a fundamental problem of the standard entropic cosmology which includes $H^2$ and $\dot{H}$ terms [Eqs.\ (\ref{eq:SmFRW01}) and (\ref{eq:SmFRW02})].
In fact, $H^2$ and $\dot{H}$ terms of entropic-force terms cannot describe a decelerating and accelerating universe predicted by the standard $\Lambda$CDM model.
This problem has been discussed in Refs. \cite{Basilakos1,Sola_2013a} and our previous paper \cite{Koma4}.
In particular, Basilakos, Polarski, and Sol\`{a} have shown that not the $H^2$ and $\dot{H}$ terms but extra constant terms are important for describing the decelerating and accelerating universe \cite{Basilakos1}. 
We emphasize that the standard entropic cosmology has such a fundamental problem, 
since entropic-force terms are usually considered to be $H^2$ and $\dot{H}$ terms.
However, a generalized black-hole entropy \cite{Tsallis2012} is expected to solve the problem, as discussed later. 
(It should be noted that the above problem does not occur in $\Lambda(t)$CDM models, as examined in Refs.\ \cite{Basilakos1,Sola_2013a}.
This is because the extra constant term is naturally obtained from an integral constant of the renormalization group equation for the vacuum energy density.
For details, see a summarized review \cite{Sola_2011b} and a recent thorough review \cite{Sola_2013b}.)

\section{Derivation of entropic-force for entropic cosmology} 
\label{Derivation of entropic-force}
In this section, in order to discuss entropic cosmology, we derive entropic-force terms from the Bekenstein entropy and a generalized black-hole entropy proposed by Tsallis and Cirto \cite{Tsallis2012}.
In Sec.\ \ref{Entropic-force from the Bekenstein entropy}, we derive the standard entropic-force term from the Bekenstein entropy, which is proportional to the surface area of a sphere with the Hubble radius.
Moreover, using an effective description for pressure, we reformulate the entropic-force model.
In Sec.\ \ref{Entropic-force from a generalized entropy}, 
we assume the generalized entropy proportional to the volume.
We derive an entropic-force term from the generalized entropy and propose a new entropic-force model. 
In the present paper, we use the Hubble radius as the preferred screen, 
because the apparent horizon coincides with the Hubble radius in the spatially flat universe \cite{Easson1}.

\subsection{Entropic-force from the Bekenstein entropy} 
\label{Entropic-force from the Bekenstein entropy}
In the standard entropic cosmology,
the modified Friedmann and acceleration equations include $H^2$ terms [Eqs.\ (\ref{eq:SmFRW01})--(\ref{eq:SmFRW02s})], as entropic-force terms.
We derive the entropic-force term from the Bekenstein entropy, according to the work of Easson {\it et al.} \cite{Easson1}.
We also reformulate the entropic-force model, using an effective pressure for entropic cosmology.

\subsubsection{Derivation of entropic-force from the Bekenstein entropy} 
\label{Derivation of entropic-force from the Bekenstein entropy} 
For entropic cosmology, we assume that the Hubble horizon has an approximate temperature $T$ and an associated entropy $S$, 
where the Hubble horizon (radius) $r_{H}$ is given by
\begin{equation}
     r_{H} = \frac{c}{H}   .
\label{eq:rH}
\end{equation}
The temperature $T$ on the Hubble horizon is given by 
\begin{equation}
 T = \frac{ \hbar H}{   2 \pi  k_{B}  } \times   \gamma  =  \frac{ \hbar }{   2 \pi  k_{B}  }  \frac{c}{ r_{H} }   \gamma    ,   
\label{eq:T0}
\end{equation}
where $k_{B}$ and $\hbar$ are the Boltzmann constant and the reduced Planck constant, respectively.
The reduced Planck constant is defined by $\hbar \equiv h/(2 \pi)$, where $h$ is the Planck constant.
The temperature considered here is obtained by multiplying the horizon temperature, $ \hbar H /( 2 \pi k_{B} ) $, by $\gamma$ \cite{Koma4}.  
In the present study, $\gamma$ is a non-negative free parameter on the order of $O(1)$, typically $\gamma \sim \frac{3}{2\pi}$ or $ \frac{1}{2}$. 
A similar parameter for the screen temperature has been discussed in Refs.\ \cite{Easson1,Cai1,Cai2,Qiu1}.
(Note that the temperature on the horizon can be evaluated as  $10^{-30}$ \textrm{K}, 
which is slightly lower than the temperature of our cosmic microwave background radiation, $2.73$ \textrm{K}.
We use the temperature on the horizon, assuming thermal equilibrium states based on a single holographic screen \cite{Easson1,Easson2}.) 

As an associated entropy on the Hubble horizon, the Bekenstein entropy $S$ is given as
\begin{equation}
 S  = \frac{ k_{B} c^3 }{  \hbar G }  \frac{A_{H}}{4}   ,
\label{eq:SH(add)}
\end{equation}
where $A_{H}$ is the surface area of the sphere with the Hubble radius $r_{H}$.
As shown in Eq.\ (\ref{eq:SH(add)}), the Bekenstein entropy is proportional to $A_{H}$.
Substituting $A_{H}=4 \pi r_{H}^2 $ into Eq.\ (\ref{eq:SH(add)}) and using $r_{H}= c/H$ as given in Eq.\ (\ref{eq:rH}), 
we obtain 
\begin{align}
S
&= \frac{ k_{B} c^3 }{  \hbar G }         \frac{A_{H}}{4}       
   =   \frac{ k_{B} c^3 }{  \hbar G }      \frac{ 4 \pi r_{H}^2 }{4}                                       
   =   \frac{ k_{B} c^3 }{ \hbar  G } \pi   \left ( \frac{c}{H} \right )^2                                     \notag \\
&=  \left ( \frac{ \pi k_{B} c^5 }{ \hbar G } \right )  \frac{1}{H^2}  =  K  \frac{1}{H^2}   ,   
\label{eq:SH(add)2}      
\end{align}
where $K$ is a positive constant \cite{Koma4} given by
\begin{equation}
  K =  \frac{  \pi  k_{B}  c^5 }{ \hbar G }     .
\label{eq:K-def}
\end{equation}
The entropic-force $F_{r}$ can be given by 
\begin{equation}
    F_{r}  =  -  \frac{dE}{dr}  =  - T \frac{dS}{dr}   \left ( =    - T \frac{dS}{dr_{H}} \right )              ,     
\label{eq:Fr}
\end{equation}
where the minus sign indicates the direction of increasing entropy or the screen corresponding to the horizon \cite{Easson1}.
Substituting Eqs.\ (\ref{eq:T0}) and (\ref{eq:SH(add)}) into Eq.\ (\ref{eq:Fr}) and using $A_{H}=4 \pi r_{H}^2$, we have the entropic-force as
\begin{align}
    F_{r}  &=    - T \frac{dS}{dr_{H}}           
              =  -  \frac{ \hbar }{ 2 \pi k_{B}  }  \frac{c}{ r_{H} }  \gamma   \times  \frac{d}{dr_{H}}  \left [ \frac{ k_{B} c^3 }{  \hbar G } \frac{4 \pi r_{H}^2 }{4}  \right ]      \notag \\         
            &=  - \gamma  \frac{c^{4}}{G}   .
\label{eq:F(add)}
\end{align}
Therefore, the pressure $p_{\rm{e_{\rm{B}}}}$ derived from the Bekenstein entropy is given by
\begin{align}
p_{\rm{e_{\rm{B}}}}  &=    \frac{ F_{r} } {A_{H}}   =  -  \gamma  \frac{c^{4}}{G}   \frac{1} {A_{H}}     =  -  \gamma  \frac{c^{4}}{G}    \frac{1} {4 \pi r_{H}^2}                                \notag \\ 
                            &=     -  \gamma  \frac{c^{4}}{G}    \frac{1} {4 \pi (c/H)^2} =  -  \gamma  \frac{c^{2}}{4 \pi G}  H^{2}                   .
\label{eq:P(add)}
\end{align}
The above derivation is based on the original idea of Easson \textit{et al}. \cite{Easson1}. 
Note that Eq.\ (\ref{eq:P(add)}) includes $\gamma$ used in this study, where $\gamma$ is a free parameter for the temperature.

In the following, in order to reformulate entropic cosmology, we consider an effective description similar to bulk viscous cosmology.
In other words, we assume an effective pressure for entropic cosmology.
We will discuss the reason for this in Sec.\ \ref{Entropic-force from a generalized entropy}.
We assume that the effective pressure $p^{\prime}$ based on the Bekenstein entropy is given by  
\begin{equation}
   p^{\prime} = p +  p_{\rm{e_{\rm{B}}}}  = p   -  \gamma  \frac{c^{2}}{4 \pi G}  H^{2}     .
\label{eq:p_eff}
\end{equation}
Using $p^{\prime}$, the acceleration equation can be written as
\begin{equation}
  \frac{ \ddot{a} }{ a }   =  -  \frac{ 4\pi G }{ 3 }  \left (  \rho +  \frac{3p^{\prime}}{c^2}  \right )   .
\label{eq:FRW2p_eff}
\end{equation}
Accordingly, substituting Eq.\ (\ref{eq:p_eff}) into Eq.\ (\ref{eq:FRW2p_eff}), we have 
\begin{align}
  \frac{ \ddot{a} }{ a }   &=  -  \frac{ 4\pi G }{ 3 }  \left (  \rho +  \frac{3 \left (  p   -  \gamma  \frac{c^{2}}{4 \pi G}  H^{2}  \right )  }{c^2}  \right )          \notag \\   
                                  &=   -  \frac{ 4\pi G }{ 3 }  \left (  \rho +  \frac{ 3p }{c^2}  \right )    +   \gamma  H^{2}    .
\label{eq:FRW2_p(add)}
\end{align}
The $ \gamma H^2$ term is the entropic-force term, which can explain the accelerated expansion of the universe.
When $\gamma =1$, Eq.\ (\ref{eq:FRW2_p(add)}) corresponds to the modified acceleration equation derived by Easson {\it et al.} \cite{Easson1}.

We now examine the continuity equation for entropic cosmology.
Using the effective pressure $p^{\prime}$, the continuity equation is expected to be given by
\begin{equation}
 \dot{\rho} + 3 \frac{\dot{a}}{a}   \left (  \rho + \frac{p^{\prime}}{c^2}  \right ) = 0 . 
\label{eq:fluid_eff}
\end{equation}
Substituting Eq.\ (\ref{eq:p_eff}) into Eq.\ (\ref{eq:fluid_eff}) and rearranging, we have  
\begin{equation}
 \dot{\rho} + 3 \frac{\dot{a}}{a} \left (  \rho + \frac{p}{c^2}  \right ) =   \gamma  \frac{3}{4 \pi G}  H^{3}    .
\label{eq:fluid_(add)}
\end{equation}
This is the modified continuity equation obtained from the effective pressure for entropic cosmology.
In the present paper, we consider Eq.\ (\ref{eq:fluid_(add)}) as the modified continuity equation for entropic cosmology derived from the Bekenstein entropy.
[Note that Eqs.\ (\ref{eq:fluid_(add)}) and (\ref{eq:fluid_thermo(add)}) are consistent with each other when $\dot{H} = - H^{2}$.
The universe for $\dot{H} = - H^{2}$ corresponds to the empty universe, as discussed in Sec.\ \ref{combined model}.]

We can determine two dimensionless constants $\alpha_{1}$ and $\beta_{1}$ included in the modified Friedmann and acceleration equations [Eqs.\ (\ref{eq:SmFRW01s}) and (\ref{eq:SmFRW02s})] using two continuity equations. 
The first continuity equation is Eq. (\ref{eq:fluid_(add)}),  
whereas the second continuity equation can be derived from the modified Friedmann and acceleration equations.
In order to obtain the second continuity equation, we use the generalized continuity equation, i.e., Eq.\ (\ref{eq:drho_General0}).
Comparing [Eqs.\ (\ref{eq:General_FRW01_f}) and (\ref{eq:General_FRW02_g})] with [Eqs.\ (\ref{eq:SmFRW01s}) and (\ref{eq:SmFRW02s})], we can set general functions as $f(t) = \alpha_{1} H^{2}$ and $g(t) = \beta_{1} H^{2}$. 
Substituting these functions into Eq.\ (\ref{eq:drho_General0}) and rearranging, we have the second continuity equation, which is given by
\begin{align}
\dot{\rho} + 3  \frac{\dot{a}}{a} \left (  \rho + \frac{p}{c^2}  \right )  
          &=  \frac{3}{4 \pi G} H \left(  - \alpha_{1} H^{2} -  \frac{ 2 \alpha_{1} H \dot{H} }{2 H }  +  \beta_{1} H^{2}      \right )     \notag \\
          &=  \frac{3}{4 \pi G} H \left(  (\beta_{1} - \alpha_{1}) H^{2} -  \alpha_{1} \dot{H} \right )    .
\label{eq:drho_2nd_(add)}
\end{align}
We expect that the two modified continuity equations, i.e., Eqs.\ (\ref{eq:fluid_(add)}) and (\ref{eq:drho_2nd_(add)}), are consistent with each other. 
Consequently, we obtain $\alpha_{1}$ and $\beta_{1}$ as 
\begin{equation}
    \alpha_{1}=0               \quad    \rm{and}    \quad    \beta_{1}=  \gamma    .
\end{equation}
Therefore, we can neglect the entropic-force term $ \alpha_{1} H^{2}$ of the modified Friedmann equation shown in Eq.\ (\ref{eq:SmFRW01s}). 
This is because, in the present paper,  we assume the effective pressure for entropic cosmology.

As mentioned previously, we consider $\gamma$ as a free parameter for the temperature. 
Moreover, a free parameter for the entropy may be required for calculating $TdS$.
However, we do not use the free parameter for the entropy, 
because we assume that the Bekenstein entropy is given by $ S  = (k_{B} c^3 A_{H}) /(4 \hbar G)$ [Eq.\ \ref{eq:SH(add)}].

\subsubsection{Entropic-force model for the Bekenstein entropy} 
\label{Entropic-force model for the Bekenstein entropy}

In Sec.\ \ref{Derivation of entropic-force from the Bekenstein entropy}, 
we derived an entropic-force term from the Bekenstein entropy and obtained an entropic-force model. 
Consequently, the modified Friedmann, acceleration, and continuity equations are summarized as 
\begin{equation}
 \left(  \frac{ \dot{a} }{ a } \right)^2    =  \frac{ 8\pi G }{ 3 } \rho  + \alpha_{1} H^2   ,  
\label{eq:SmFRW01s_summ}
\end{equation}
\begin{equation}
\frac{ \ddot{a} }{ a }  =      -  \frac{ 4\pi G }{ 3 } \left ( \rho +  \frac{3p}{c^2} \right )        +  \beta_{1}  H^2  ,
\label{eq:SmFRW02s_summ}
\end{equation}
and
\begin{equation}
 \dot{\rho} + 3 \frac{\dot{a}}{a} \left (  \rho + \frac{p}{c^2}  \right ) =   \gamma  \frac{3}{4 \pi G}  H^{3}    , 
\label{eq:fluid_(add)summ}
\end{equation}
where $\alpha_{1}$ and $\beta_{1}$ are given by 
\begin{equation}
    \alpha_{1}=0               \quad    \rm{and}    \quad    \beta_{1}=  \gamma    .
\label{eq:alpha_summ}
\end{equation}
Note that we leave $\alpha_{1} H^2$ in Eq.\ (\ref{eq:SmFRW01s_summ}) in order to discuss a combined model later.

\subsection{Entropic-force from a generalized entropy} 
\label{Entropic-force from a generalized entropy}

Thus far, we have considered the Bekenstein entropy to discuss the standard entropic cosmology.
Instead of the Bekenstein entropy, we now assume a generalized black-hole entropy proposed by Tsallis and Cirto \cite{Tsallis2012}.

\subsubsection{Derivation of entropic-force from a generalized entropy} 
\label{Derivation of entropic-force from a generalized entropy}

Recently, Tsallis and Cirto \cite{Tsallis2012} examined a black-hole entropy using appropriate nonadditive generalizations for $d$-dimensional systems and suggested a generalized black-hole entropy.
In the following, we introduce the generalized black-hole entropy, according to the work of Tsallis and Cirto \cite{Tsallis2012}.
In their study, a nonadditive entropy (for a set of $W$ discrete states) is defined by
\begin{equation}
S_{\delta}  =  k_{B} \sum_{i=1}^{W}  p_{i} \left ( \ln \frac{1}{{p_{i}}  } \right ) ^{\delta}  \quad (\delta >0)   ,
\end{equation}
where $p_{i}$ is a probability distribution \cite{Tsa1,Tsallis2012}. 
When $\delta = 1$, $S_{\delta}$ recovers the Boltzmann--Gibbs entropy given by
\begin{equation}
 S_{\rm{BG}} = - k_{B} \sum_{i=1}^{W}  p_{i} \ln p_{i}    .
\end{equation}
If we compose two probabilistically independent subsystems $A$ and $B$, 
the Boltzmann--Gibbs entropy $S_{\rm{BG}}$ is additive:
\begin{equation}
 S_{\rm{BG}}(A+B)  =  S_{\rm{BG}}(A)  + S_{\rm{BG}}(B)  .
\end{equation}
However, when $\delta \neq 1$, $S_{\delta}$ is nonadditive: 
\begin{equation}
 S_{\delta}(A+B)  \neq  S_{\delta}(A)  + S_{\delta}(B)  .
\end{equation}
This is because, for $\delta>0$, $S_{\delta}(A+B)$ is given by   
\begin{equation}
 \frac{ S_{\delta}(A+B) } { k_{B} } =  \left (  \left [  \frac{ S_{\delta}(A)  } { k_{B} } \right ] ^{1/\delta}    + \left [  \frac{ S_{\delta}(B)  } { k_{B} } \right ] ^{1/\delta}    \right ) ^{\delta} .
\end{equation}
In Ref.\ \cite{Tsallis2012}, Tsallis and Cirto demonstrated that a generalized black-hole entropy can be written as
\begin{equation}
 \frac{ S_{\delta = 3/2 } } { k_{B} }  \propto \left (  \frac{ S^{\prime}  } { k_{B} } \right )   ^{\frac{3}{2}}   , 
\label{eq:S-SBH}
\end{equation}
where the Bekenstein entropy $ S^{\prime} $ is given by
\begin{equation}
 S^{\prime}  = \frac{ k_{B} c^3 }{  \hbar G }  \frac{A_{H}^{\prime}}{4}    ,
\label{eq:SBH}
\end{equation}
and $A_{H}^{\prime}$ is the event horizon area of a black hole.
For details, see Ref.\ \cite{Tsallis2012}.

We now apply the generalized black-hole entropy to an entropy for entropic cosmology. 
To this end, substituting Eq.\ (\ref{eq:SBH}) into Eq.\ (\ref{eq:S-SBH}), 
replacing $A_{H}^{\prime}$ with $A_{H}$, and rearranging, the entropy $S$ on the Hubble horizon can be evaluated as 
\begin{equation}
  S  \propto   A_{H}^{\frac{3}{2}}  \propto   r_{H}^{3}  ,
\label{eq:Sprop}
\end{equation}
where $A_{H}$ is the surface area of the sphere with the Hubble radius $r_{H}$, and, therefore, $A_{H}=4 \pi r_{H}^2$.
Because of nonadditive generalizations, the generalized entropy is proportional to $r_{H}^{3}$, i.e., volume. 
Accordingly, from Eqs.\ (\ref{eq:SBH}) and (\ref{eq:Sprop}), we assume a generalized entropy $S_{g}$ given by 
\begin{equation}
S_{g}  =    \frac{  \pi  k_{B} c^3 }{  \hbar G } \times  \zeta r_{H}^{3}           ,
\label{eq:S(nonadd)}
\end{equation}
where $\zeta$ is a non-negative free-parameter.
Note that $\zeta$ is a dimensional constant. We hereinafter refer to $S_{g}$ as the generalized entropy on the Hubble horizon.
Substituting $r_{H}= c/H$ [Eq.\ (\ref{eq:rH})] into Eq.\ (\ref{eq:S(nonadd)}), 
we obtain
\begin{align}
S_{g} 
&= \frac{  \pi  k_{B} c^3 }{  \hbar G } \times  \zeta r_{H}^{3}
   =   \frac{  \pi  k_{B} c^3 }{  \hbar G } \times  \zeta  \left ( \frac{c}{H} \right )^{3} \notag \\
&=  \left ( \frac{ \pi k_{B} c^5 }{ \hbar G } \right ) c \zeta  \frac{1}{H^3}  =  K  c \zeta  \frac{1}{H^3}   ,  
\label{eq:S(nonadd)2}      
\end{align}
where $K$ is $ \pi  k_{B}  c^5  / (\hbar G) $ [Eq.\ (\ref{eq:K-def})].

In order to calculate an entropic-force, we use the generalized entropy [Eq.\ (\ref{eq:S(nonadd)})] and the temperature $T$ on the horizon [Eq.\ (\ref{eq:T0})].
Substituting Eqs.\ (\ref{eq:T0}) and (\ref{eq:S(nonadd)}) into Eq.\ (\ref{eq:Fr}) and using $r_{H}= c/H$, we obtain the entropic-force as
\begin{align}
    F_{r}  &=  - T \frac{dS}{dr}     =    - T \frac{dS_{g}}{dr_{H}}            \notag \\
           &=  -  \frac{ \hbar }{ 2 \pi k_{B}  }  \frac{c}{ r_{H} }  \gamma   \times   \frac{d}{dr_{H}}  \left [ \frac{  \pi  k_{B} c^3 }{  \hbar G } \times  \zeta r_{H}^{3}    \right ]     \notag \\
           &=  -  \gamma  \frac{c^{4}}{G} \left ( \frac{3}{2} \zeta r_{H} \right ) =  -  \gamma  \frac{c^{4}}{G} \left ( \frac{3 c \zeta}{2} \frac{1}{H} \right )    .
\label{eq:F(nonadd)}
\end{align}
Therefore, the pressure $p_{\rm{e}_g}$ derived from the generalized entropy is given as 
\begin{align}
   p_{\rm{e}_g}  &=    \frac{ F_{r} } {A_{H}}   =  -  \gamma  \frac{c^{4}}{G} \left ( \frac{3 c \zeta}{2} \frac{1}{H} \right )   \frac{1} {4 \pi r_{H}^2}     \notag \\ 
                     &=  -  \gamma  \frac{c^{4}}{G} \left ( \frac{3 c \zeta}{2} \frac{1}{H} \right )   \frac{1}{4 \pi (c/H)^2}                                  
                       =  -  \gamma  \frac{c^{2}}{4 \pi G}   \frac{3 c \zeta }{2} H                         .
\label{eq:P(nonadd)}
\end{align}
In this equation, $3c \zeta/2$ is shown separately in order to clarify the difference between Eqs.\ (\ref{eq:P(add)}) and (\ref{eq:P(nonadd)}). 
Similarly, $3c \zeta/2$ is shown separately, in the following.

In the present study, we assume an effective pressure $p^{\prime}$ for entropic cosmology.
Using Eq.\ (\ref{eq:P(nonadd)}), the effective pressure $p^{\prime}$ based on the generalized entropy is given by  
\begin{equation}
   p^{\prime} = p +  p_{\rm{e}_g}  = p     -  \gamma  \frac{c^{2}}{4 \pi G}  \frac{3 c \zeta }{2} H     .
\label{eq:p_eff(nonadd)}
\end{equation}
Substituting Eq.\ (\ref{eq:p_eff(nonadd)}) into Eq.\ (\ref{eq:FRW2p_eff}), we have 
\begin{align}
  \frac{ \ddot{a} }{ a }   &=  -  \frac{ 4\pi G }{ 3 }  \left (  \rho +  \frac{3 \left (  p   -   \gamma  \frac{c^{2}}{4 \pi G}  \frac{3 c \zeta }{2} H    \right )  }{c^2}  \right )          \notag \\   
                                  &=   -  \frac{ 4\pi G }{ 3 }  \left (  \rho +  \frac{ 3p }{c^2}  \right )    +   \gamma  \frac{3 c \zeta }{2} H      .
\label{eq:FRW2_p(nonadd)}
\end{align}
This equation is the modified acceleration equation derived from the generalized entropy. 
The $ H$ term on the right-hand side corresponds to an extra driving term to explain the accelerated expansion of the universe. 
The extra driving term is found not to be the $H^{2}$ term but rather to be the $H$ term, unlike in the case of the standard entropic cosmology.
The continuity equation is expected to be given by 
\begin{equation}
 \dot{\rho} + 3 \frac{\dot{a}}{a}   \left (  \rho + \frac{p^{\prime}}{c^2}  \right ) = 0 . 
\label{eq:fluid_eff2}
\end{equation}
Substituting Eq.\ (\ref{eq:p_eff(nonadd)}) into Eq.\ (\ref{eq:fluid_eff2}) and rearranging, we have  
\begin{equation}
 \dot{\rho} + 3 \frac{\dot{a}}{a} \left (  \rho + \frac{p}{c^2}  \right ) =    \gamma  \frac{3}{4 \pi G}  \frac{3 c \zeta }{2} H^{2}    .
\label{eq:fluid_(nonadd)}
\end{equation}
In this paper, we consider Eq.\ (\ref{eq:fluid_(nonadd)}) as the modified continuity equation based on the generalized entropy  \cite{C_21}.

Interestingly, the effective pressure, Eq.\ (\ref{eq:p_eff(nonadd)}), is similar to Eq.\ (\ref{eq:Bulk_p}), i.e., $p^{\prime} = p - 3 \eta H$, for bulk viscous cosmology.
Therefore, Eqs.\ (\ref{eq:FRW2_p(nonadd)}) and (\ref{eq:fluid_(nonadd)}) are also similar to Eqs.\ (\ref{eq:Bulk_FRW2}) and (\ref{eq:Bulk_fluid1}) for bulk viscous cosmology.
This is probably because the generalized entropy assumed here is proportional not to the surface area but to the volume.
However, the similarity may be interpreted as a sign that the generalized entropy behaves as if it were a classical entropy generated by bulk viscous stresses.
In other words, the generalized entropy may be related to a bulk viscosity of cosmological fluids through a holographic screen.
In fact, the bulk viscosity is usually the only thing that can generate an entropy in the homogeneous and isotropic universe.
Accordingly, this interpretation may help to explain the origin of the bulk viscosity of cosmological fluids.
As an alternative interpretation, the bulk viscosity may be derived from an extra entropy that is proportional to the volume of the universe.

\subsubsection{Entropic-force model for a generalized entropy} 
\label{Entropic-force model for a generalized entropy}

In Sec.\ \ref{Derivation of entropic-force from a generalized entropy}, $H$ terms are derived from a generalized entropy on the horizon. 
Accordingly, we expect that the modified Friedmann and acceleration equations are summarized, respectively, as 
\begin{equation}
  \left(  \frac{ \dot{a} }{ a } \right)^2   =  \frac{ 8\pi G }{ 3 } \rho  +  \hat{\alpha}_{3}  H ,  
\label{eq:mFRW01(nonadd)}
\end{equation}
and 
\begin{equation}
  \frac{ \ddot{a} }{ a } =    -  \frac{ 4\pi G }{ 3 }  \left (  \rho +  \frac{3p}{c^2}  \right )   +  \hat{\beta}_{3} H    , 
\label{eq:mFRW02(nonadd)}
\end{equation}
where $\hat{\alpha}_{3}$ and  $\hat{\beta}_{3}$ are defined by  
\begin{equation}
  \hat{\alpha}_{3}   \equiv  \alpha_{3}  H_{0}      \quad     \textrm{and}    \quad     \hat{\beta}_{3} \equiv  \beta_{3}  H_{0}       .
\label{eq:ab3-H0}
\end{equation}
The two coefficients $\alpha_3$ and $\beta_3$ are dimensionless constants, and $H_{0}$ is the present value of the Hubble parameter.
As shown in Eq.\ (\ref{eq:fluid_(nonadd)}), the modified continuity equation is given by
\begin{equation}
 \dot{\rho} + 3 \frac{\dot{a}}{a} \left (  \rho + \frac{p}{c^2}  \right ) =    \gamma  \frac{3}{4 \pi G}  \frac{3 c \zeta }{2} H^{2}    .
\label{eq:fluid_(nonadd)2}
\end{equation}

We can determine $\hat{\alpha}_{3}$ and $\hat{\beta}_{3}$, which are included in Eqs.\ (\ref{eq:mFRW01(nonadd)}) and (\ref{eq:mFRW02(nonadd)}), from two continuity equations.  
In this case, the first continuity equation is Eq.\ (\ref{eq:fluid_(nonadd)2}). 
The second continuity equation (derived from the modified Friedmann and acceleration equations) is calculated from the generalized continuity equation, i.e., Eq.\ (\ref{eq:drho_General0}). 
Comparing [Eqs.\ (\ref{eq:General_FRW01_f}) and (\ref{eq:General_FRW02_g})] with [Eqs.\ (\ref{eq:mFRW01(nonadd)}) and (\ref{eq:mFRW02(nonadd)})], 
we can set general functions as $f(t) = \hat{\alpha}_{3}  H$ and $g(t) =  \hat{\beta}_{3}  H$.
Therefore, substituting these functions into Eq.\ (\ref{eq:drho_General0}) and rearranging, we have the second continuity equation given by
\begin{align}
\dot{\rho} + 3  \frac{\dot{a}}{a} \left (  \rho + \frac{p}{c^2}  \right )  
          &=  \frac{3}{4 \pi G} H \left(  - \hat{\alpha}_{3} H -  \frac{  \hat{\alpha}_{3} \dot{H} }{2 H }  +  \hat{\beta}_{3}  H      \right )     \notag \\
          &=  \frac{3}{4 \pi G} H \left(  (\hat{\beta}_{3} - \hat{\alpha}_{3}) H - \hat{\alpha}_{3} \frac{  \dot{H} }{2 H }         \right )    .
\label{eq:drho_2nd}
\end{align}
Since Eqs.\ (\ref{eq:fluid_(nonadd)2}) and (\ref{eq:drho_2nd}) are expected to be consistent with each other,  
we obtain $\hat{\alpha}_{3}$ and $\hat{\beta}_{3}$ as 
\begin{equation}
    \hat{\alpha}_{3} =0               \quad    \textrm{and}    \quad    \hat{\beta}_{3} = \gamma  \frac{3c \zeta}{2}   .
\end{equation}
Accordingly, the dimensionless constants $\alpha_{3}$ and $\beta_{3}$ are given as 
\begin{equation}
    \alpha_{3} =0    \quad   \textrm{and}    \quad    \beta_{3} = \frac{\hat{\beta}_{3}}{H_{0}} =  \gamma  \frac{3c \zeta}{2 H_{0}}   . 
\label{eq:alpha3_beta3_s}
\end{equation}
As expected, we can neglect the entropic-force term $\hat{\alpha}_{3} H$ of the modified Friedmann equation, Eq.\ (\ref{eq:mFRW01(nonadd)}).
This is because, in the above discussion and in the present study, we assume an effective pressure for entropic cosmology.
Note that we leave $\hat{\alpha}_{3} H$ in Eq.\ (\ref{eq:mFRW01(nonadd)}) in order to discuss a combined model in the next section.

\begin{table}[t]
\caption{Entropic-force terms $g_{\textrm{e}}(t)$ for the modified acceleration equation.
The Bekenstein and Generalized columns indicate the information for the entropic-force models derived from the Bekenstein and generalized entropies, respectively.}
\label{tab-summary}
\newcommand{\m}{\hphantom{$-$}}
\newcommand{\cc}[1]{\multicolumn{1}{c}{#1}}
\renewcommand{\tabcolsep}{1.5pc} 
\renewcommand{\arraystretch}{1.5} 
\begin{tabular}{@{}lllll}
\hline
\hline
$\textrm{Parameter}$           &   $\textrm{Bekenstein}$         &   $\textrm{Generalized}$   \\
\hline
$\rm{Entropy}$                                      &  $ \frac{ \pi k_{B} c^3 }{  \hbar G } \times  r_{H}^2   $     & $\frac{  \pi  k_{B} c^3 }{  \hbar G } \times  \zeta r_{H}^{3}   $   \\
$g_{\textrm{e}}(t)$                    &  $\beta_{1}  H^{2}$                                                       & $\hat{\beta}_{3} H$   \\
\hline
\hline
\end{tabular}\\
 \end{table}

In this section, we derive entropic-force terms from the Bekenstein and generalized entropies, 
assuming that the temperature on the horizon is given by Eq.\ (\ref{eq:T0}), i.e., $ T = \gamma \hbar H /( 2 \pi k_{B} ) $.
Consequently, $H^{2}$ terms are derived from the Bekenstein entropy, 
whereas $H$ terms are derived from the generalized entropy, which is proportional to volume.
The two entropic-force terms for the modified acceleration equation are summarized in Table \ref{tab-summary}. 
Interestingly, the $H$ term is similar to an extra driving term for bulk viscous cosmology.
Therefore, we assume an effective pressure for entropic cosmology.
The modified acceleration equation is found to include the entropic-force terms, whereas the Friedmann equation does not include the entropic-force terms. 
Based on these results, we propose two entropic-force models.
In the next section, we discuss solutions of the two entropic-force models.

\section{Combined model}
\label{combined model}
In the previous section, the Friedmann equation was found not to include entropic-force terms, 
because we use an effective pressure for entropic cosmology.
Moreover, two entropic-force terms, i.e., $H^{2}$ and $H$ terms, are discussed separately, 
as shown in Secs.\ \ref{Entropic-force from the Bekenstein entropy} and \ref{Entropic-force from a generalized entropy}.
This is because the $H^{2}$ term is derived from the Bekenstein entropy, 
whereas the $H$ term is derived from a generalized entropy.
However, in this section, in order to obtain general solutions that are widely used, we consider a combined model in which the Friedmann and acceleration equations include both $H^{2}$ and $H$ terms as the entropic-force terms. 
In order to solve the equations, we extend our solution method, which is discussed in Ref.\ \cite{Koma4}.
Note that the two extra driving terms for the acceleration equation, i.e., $H^{2}$ and $H$ terms, have been examined in bulk viscous cosmology. For example, see the work of Avelino and Nucamendi \cite{Avelino2}.  
Of course, similar extra driving terms have been discussed using a variable cosmological term. 
For instance, Basilakos, Plionis, and Sol\`{a} \cite{Sola_2009} have examined $H^{2}$ and $H$ terms in detail.
Several types of variable cosmological terms, which include constant terms, are closely investigated in Ref. \cite{Sola_2009}.  
For variable $\Lambda$ cosmologies, see Ref. \cite{Overduin1}, a recent review \cite{Sola_2013b}, and the references therein.
We point out that cosmological equations and their solutions discussed in this section are similar to those examined in the above previous works.

\subsection{Formulations for the combined model}
\label{Formulations for the combined model}
For a combined model, we consider both $H^{2}$ and $H$ terms as extra driving terms.
Accordingly, the modified Friedmann equation is given as 
\begin{equation}
  \left(  \frac{ \dot{a} }{ a } \right)^2   =  H^{2}  = \frac{ 8\pi G }{ 3 } \rho  + \alpha_{1} H^{2} + \hat{\alpha}_{3}  H ,  
\label{eq:FRW01(g)}
\end{equation}
and the modified acceleration equation is given as
\begin{align}
  \frac{ \ddot{a} }{ a } 
                     &=    \dot{H} +H^{2}                                                                                               \notag \\
                     &=    -  \frac{ 4\pi G }{ 3 } (  1+  3 w  ) \rho  +  \beta_{1} H^{2} +  \hat{\beta}_{3} H    , 
\label{eq:FRW02(g)}
\end{align}
where $w$ is given by
\begin{equation}
  w = \frac{ p } { \rho  c^2 }.
\label{eq:w_(2)}
\end{equation}
$\hat{\alpha}_{3}$ and $\hat{\beta}_{3}$ are defined by  
\begin{equation}
  \hat{\alpha}_{3}   \equiv  \alpha_{3}  H_{0}      \quad     \textrm{and}    \quad     \hat{\beta}_{3} \equiv  \beta_{3} H_{0},
\label{eq:ab3-H0_(2)}
\end{equation}
and $w$ represents the equation of state parameter for a generic component of matter.
For non-relativistic matter (or the matter-dominated universe) $w$ is $0$, and for relativistic matter (or the radiation-dominated universe) $w$ is $1/3$.
The four coefficients $\alpha_1$, $\beta_1$, $\alpha_3$, and $\beta_3$ are dimensionless constants.
For entropic-force models for the Bekenstein and generalized entropies, three dimensionless constants are set as $\alpha_{1} = \alpha_{3} = \beta_{3} =0$, and $\alpha_{1} = \alpha_{3} = \beta_{1} =0$, respectively. 

Coupling [$(1+3w) \times $ Eq.\ (\ref{eq:FRW01(g)})] with [$2 \times $ Eq.\ (\ref{eq:FRW02(g)})] and rearranging, we obtain  
\begin{equation}
 \dot{H}  = \frac{ dH }{ dt }  = -  C_{1} H^2    +   \hat{C}_{3} H, 
\label{eq:dHC1C3h}
\end{equation}
where
\begin{equation}
  C_{1} = \frac{  3(1 + w)  -  \alpha_{1}(1+3w) - 2\beta_{1}  }{   2   } ,   
\label{eq:C1}
\end{equation}
and 
\begin{equation}
  \hat{C}_{3} = \frac{  \hat{\alpha}_{3}(1+3w) + 2\hat{\beta}_{3}  }{   2   } .  
\label{eq:C3h}
\end{equation}
Equation (\ref{eq:dHC1C3h}) includes not only the $H^{2}$ term but also the $H$ term.
(In our previous study \cite{Koma4}, the $H$ term was not considered.)
Here, we point out that $C_{1}$ is a dimensionless parameter, whereas $\hat{C}_{3}$ is a dimensional parameter.
Substituting Eq.\ (\ref{eq:ab3-H0_(2)}) into Eq.\ (\ref{eq:C3h}), 
we obtain a dimensionless parameter $C_{3}$ given by
\begin{equation}
  {C}_{3} = \frac{ \hat{C}_{3} }{ H_{0} } = \frac{ \alpha_{3}(1+3w) + 2 \beta_{3}  }{   2   } .  
\label{eq:C3}
\end{equation}

From Eq.\ (\ref{eq:dHC1C3h}), $(dH/da) a$ is calculated as
\begin{align}
\left ( \frac{dH}{da} \right )  a     &=       \left ( \frac{dH}{dt} \right )   \frac{dt}{da} a  
                                                   =    ( - C_{1} H^2 + \hat{C}_{3} H)   \frac{a}{\dot{a}}                      \notag \\
                                                 &= (-  C_{1} H^2 + \hat{C}_{3} H ) \frac{1}{H}                  
                                                  =  - C_{1}  H  + \hat{C}_{3}  . 
\label{eq:dHda_C1HC3h}
\end{align}
We can rearrange Eq.\ (\ref{eq:dHda_C1HC3h}) as
\begin{equation}
  \frac{ d H }{ d N } =  - C_{1} H  + \hat{C}_{3} ,
\label{eq:dHdN}
\end{equation}
where $N$ is defined by 
\begin{equation}
   N  \equiv \ln a   \quad \textrm{and therefore} \quad  dN   = \frac{da}{a}   .  
\end{equation} 
Because of the $\hat{C}_{3}$ term, Eq.\ (\ref{eq:dHdN}) is slightly more complicated than the equation examined in Ref.\ \cite{Koma4}.  
We can solve Eq.\ (\ref{eq:dHdN}), as discussed in the next subsection.

\subsection{Solutions for the combined model in the single-fluid-dominated universe}
\label{Solutions for the combined model in the single-fluid-dominated universe}

In the previous subsection, we obtain Eq.\ (\ref{eq:dHdN}) for the combined model.
We can solve Eq.\ (\ref{eq:dHdN}) when $C_{1}$ and $\hat{C}_{3}$ are constant.
In fact, $C_{1}$ and $\hat{C}_{3}$ are constant when five parameters, i.e., $\alpha_{1}$, $\beta_{1}$, $\hat{\alpha}_{3}$, $\hat{\beta}_{3}$, and $w$, are constant.
Therefore, we assume that these five parameters are constant.
This indicates that we assume a single-fluid-dominated universe. 
For example, $w$ is $0$ for the matter-dominated universe and $1/3$ for the radiation-dominated universe. In the following, we consider $C_{1}$, $\hat{C_{3}}$, and $C_{3}$ as non-negative free parameters.  For simplicity, we also assume $C_{1}  > 0$.

When $C_1$ and $\hat{C}_{3}$ are constant, Eq.\ (\ref{eq:dHdN}) is integrated as
\begin{equation}
 \int \frac{dH}{- C_{1} H + \hat{C}_{3} } = \int dN   .  
\end{equation}
Solving this integral, and using $N = \ln a$, we have 
\begin{equation}
C_{1} H - \hat{C}_{3}  =  D a^{- C_{1}}      , 
\end{equation}
and dividing this equation by $C_{1}$ gives 
\begin{equation}
   H - \frac{\hat{C}_{3}}{C_{1}}  = \frac{D}{C_{1}} a^{- C_{1}}      , 
\label{eq:Solve1}
\end{equation}
where $D$ is an integral constant.
Dividing Eq.\ (\ref{eq:Solve1}) by $ H_{0} - (\hat{C}_{3}/C_{1})  =  (D/C_{1}) a_{0}^{- C_{1}}$, we have   
\begin{equation}
 \frac{ H - (\hat{C}_{3}/C_{1})  }{ H_{0} - (\hat{C}_{3}/C_{1})  }   =  \left ( \frac{ a } {  a_{0} } \right )^{ -C_{1}}   ,
\label{eq:H/H0}
\end{equation}
where $a_0$ is the present value of the scale factor.
Rearranging Eq.\ (\ref{eq:H/H0}), and substituting $C_{3}= \hat{C}_{3}/H_{0}$ [Eq.\ (\ref{eq:C3})] into the resulting equation, we obtain 
\begin{align}
 \frac{H} {H_{0}} &=  \left ( 1-  \frac{1}{H_{0}} \frac{ \hat{C}_{3} }{ C_{1} }  \right )   \left ( \frac{ a } {  a_{0} } \right )^{ -C_{1}}   +  \frac{1}{H_{0}} \frac{\hat{C}_{3}}{C_{1}}    \notag\\
                       &=  \left ( 1-  \frac{ C_{3} }{ C_{1} }  \right )   \left ( \frac{ a } {  a_{0} } \right )^{ -C_{1}}   +  \frac{ C_{3} }{ C_{1} }    .
\label{eq:H/H0_(2)}
\end{align}
Equation (\ref{eq:H/H0_(2)}) indicates that $C_{1}$ and $C_{3}$ play an important role in the combined model.
We can determine $C_{1}$ and $C_{3}$ from Eqs.\ (\ref{eq:C1}) and (\ref{eq:C3}), respectively.
Since $C_{3}$ is related to the $H$ terms, Eq.\ (\ref{eq:H/H0_(2)}) is somewhat complicated.
For the case in which $C_{3}=0$, typical results have been discussed in our previous study \cite{Koma4}. 
For example, when $C_{3}=0$, $H/H_{0}$ for $C_{1} = 2$, $1.5$, $1$, and $0$ are consistent with $H/H_{0}$ for the radiation-dominated, matter-dominated, empty, and $\Lambda$-dominated universes, respectively \cite{Koma4}.

\subsubsection{Scale factor $a$} 
\label{Scale factor} 

We examine the scale factor $a(t)$, using Eq.\ (\ref{eq:H/H0_(2)}).
For this purpose, Eq.\ (\ref{eq:H/H0_(2)}) is rearranged as 
\begin{equation}
 \tilde{ H }    =   ( 1- A) \tilde{a}^{-C_{1}}   +  A    ,
\label{eq:H/H0_(2)AB}
\end{equation}
where $\tilde{ H }$,  $\tilde{ a }$, and $A$ are defined by 
\begin{equation}
 \tilde{ H } \equiv \frac{H}{H_0} ,  \quad  \quad  \tilde{ a } \equiv \frac{a}{a_0} ,  \quad   \quad   A \equiv  \frac{ C_{3} }{ C_{1} } .
\label{eq:def_H_a}
\end{equation}
Multiplying Eq.\ (\ref{eq:H/H0_(2)AB}) by $\tilde{ a } $, we obtain
\begin{equation}
\tilde{H} \tilde{ a }   =   \tilde{a} [( 1- A) \tilde{a}^{-C_{1}}  +  A]      .
\label{eq:dadt}
\end{equation}
On the other hand, we can calculate $\tilde{H} \tilde{ a }$ as  
\begin{equation}
\tilde{H} \tilde{ a }    =  \frac{H}{ H_{0} }  \frac{ a }{ a_0}  
                                 = \frac{ \dot{a}/a }{ H_{0} }  \frac{ a }{ a_0}  
                                 = \frac{ \dot{a}  }{ H_{0} a_{0} }                                                                       
                                 = \frac{  a_{0} \frac{d}{dt}  \left ( \frac{a}{a_0} \right )  }{  H_{0} a_{0}  }
                               = \frac{1}{ H_{0} }  \frac{d \tilde{a} }{dt}     .
\label{eq:Ha_tilde}
\end{equation}
Therefore, substituting Eq.\ (\ref{eq:Ha_tilde}) into Eq.\ (\ref{eq:dadt}), we have 
\begin{equation}
\frac{1}{ H_{0} } \frac{ d\tilde{a} } { dt }      =   \tilde{a} [( 1- A) \tilde{a}^{-C_{1}}  +  A]  .
\label{eq:dadt_2}
\end{equation}
As examined in Ref.\ \cite{Avelino2}, integrating Eq.\ (\ref{eq:dadt_2}), we obtain

\begin{equation}
\int ^{\tilde{a}} _{1}  \frac{ d a^{\prime}  }{  a^{\prime} [( 1- A) a^{\prime  -C_{1}}  +  A]  }
= \int ^{t}_{t_{0}} H_{0} dt^{\prime} =H_{0} (t -t_{0})   , 
\label{eq:int_a-t}
\end{equation}
where $t_{0}$ represents the present time. 
Solving this integral yields 
\begin{equation}
 \frac{1}{A C_{1}} \ln \left [  1 - A + A \tilde{a}^{C_{1} }  \right ]  =   H_{0} (t -t_{0})   . 
\label{eq:2AB}
\end{equation}
Moreover, solving Eq.\ (\ref{eq:2AB}) for $\tilde{a}$ and substituting Eq.\ (\ref{eq:def_H_a}) into the result, we finally have   
\begin{equation} 
   \frac{a}{a_{0}}  = \left [   1    +  \frac{C_{1}}{C_{3}}  \Bigl ( \exp[ C_{3} H_{0}( t - t_{0} ) ]  -1  \Bigl  )   \right  ]^{\frac{1}{C_{1}}}     ,
\label{eq:a-t}
\end{equation}
where $C_{1} > 0$ and $C_{3} > 0$ are assumed.
Avelino and Nucamendi discussed this type of equation for bulk viscous cosmology in detail \cite{Avelino2}.
Note that, unlike the combined model of the present study, the Friedmann equation for bulk viscous cosmology does not have extra terms.
When $C_{3} =0$, the normalized scale factor $a/a_{0}$ is given by 
\begin{equation} 
   \frac{a}{a_{0}}  =    \begin{cases}
                            (C_{1} H_{0} t )^{  \frac{1}{C_{1}} }  &   (C_{1} \neq 0) ,        \\
                            \exp[ H_{0}( t - t_{0} ) ]                &   (C_{1}      = 0) ,         \\
\end{cases}
\label{eq:a-t(C3=0)}
\end{equation}
where $t_{0}$ is set to be $1/(C_{1} H_{0})$ \cite{Koma4}.

We now examine the deceleration parameter $q_{0}$, which is used to discuss the expansion of the universe \cite{Weinberg1,Ryden1,Roy1}.
The deceleration parameter is defined by
\begin{equation}
q_{0} \equiv  - \left ( \frac{\ddot{a} } {a H^{2}} \right )_{t=t_0}  .
\label{eq:q_def}
\end{equation}
Substituting Eq.\ (\ref{eq:dHC1C3h}) into Eq.\ (\ref{eq:FRW02(g)}), we obtain
\begin{equation}
  \frac{ \ddot{a} }{ a }  =  \dot{H} + H^{2}   =  (-  C_{1} H^2    +   \hat{C}_{3} H)   + H^{2} = (1- C_{1}) H^2   +   \hat{C}_{3} H, 
\end{equation}
and rearranging the result yields
\begin{equation}
  - \frac{ \ddot{a} }{ a H^2 }  =   C_{1}    -  \frac{ \hat{C}_{3} }{ H } - 1 . 
\label{eq:ddota_a_C1C3h}
\end{equation}
This equation is a temporal deceleration parameter $q(t)$.
Even if $C_{1}$ and $\hat{C}_{3}$ are constant, $q(t)$ can vary during the evolution of the universe depending on the $ \hat{C}_{3} / H$ term. 
Substituting Eq.\ (\ref{eq:ddota_a_C1C3h}) into Eq.\ (\ref{eq:q_def}) and using $\hat{C}_{3} =C_{3} H_{0}$ [Eq.\ (\ref{eq:C3})], 
the deceleration parameter $q_{0}$ is given as 
\begin{equation}
q_{0} =  C_{1} -1  -  \frac{ \hat{C}_{3} }{ H_{0} }  = C_{1}  -  C_{3} -1 . 
\label{eq:q0C1C3}
\end{equation}
When $C_{1}- C_{3} <1 $, $q_{0}$ is negative corresponding to a positive acceleration.
This will be discussed in Sec.\ \ref{Comparison}.
Note that, in calculating $q_{0}$, we do not assume the single-fluid-dominated universe.

The temporal deceleration parameter $q(t)$ [Eq. (\ref{eq:ddota_a_C1C3h})] is a negative constant when $C_{1}<1$ and $\hat{C}_{3} =0$.
In other words, the expansion of the universe uniformly accelerates when $C_{1}<1$ and $\hat{C}_{3} =0$.  
Therefore, entropic-force models for the Bekenstein entropy cannot describe a decelerating and accelerating universe. 
As discussed in Sec. \ref{Standard entropic-force models}, a similar fact has been examined in Refs.\ \cite{Koma4,Basilakos1,Sola_2013a}, 
based on the standard entropic cosmology which includes $H^{2}$ and $\dot{H}$ terms [Eqs. (\ref{eq:SmFRW01}) and (\ref{eq:SmFRW02})].

\subsubsection{Luminosity distance $d_{L}$} 
\label{Luminosity distance}

The luminosity distance is an important parameter for investigating the accelerated expansion of the universe. 
Therefore, we examine the luminosity distance $d_{L}$ of the single-fluid-dominated universe in the combined model.
The luminosity distance \cite{Sato1} is generally given as
\begin{equation}
 d_{L}(z)   = \frac{ c (1+z) }{ H_{0} }  \int_{1}^{1+z}  \frac{dy} {F(y)} ,
\label{eqA:dL-def1}  
\end{equation}
where the integrating variable $y$ and the function $F(y)$ are given by 
\begin{equation}
  y = \frac{a_0} {a}  , 
\end{equation}
and
\begin{equation}
F(y)   = \frac{ H }{ H_{0} } ,
\label{eqA:dL-def2}  
\end{equation}
and $z$ is the redshift defined by 
\begin{equation}
 1 + z \equiv  y = \frac{ a_0 }{ a } .
\label{eqA:z-def}  
\end{equation}
For the combined model, substituting Eq.\ (\ref{eq:H/H0_(2)}) into Eq.\ (\ref{eqA:dL-def2}), and using $y = a_{0}/a$, we obtain $F(y)$ as
\begin{align}
F(y) = \frac{ H }{ H_{0} }  
         & =  \left ( 1-  \frac{ C_{3} }{ C_{1} }  \right )   \left ( \frac{ a } {  a_{0} } \right )^{ -C_{1}}   +  \frac{ C_{3} }{ C_{1} }   \notag \\
         &=  \left ( 1-  \frac{ C_{3} }{ C_{1} }  \right )   y^{ C_{1}}   +  \frac{ C_{3} }{ C_{1} }   .
\label{eqA:FyC1C3}  
\end{align}
Accordingly, substituting Eq.\ (\ref{eqA:FyC1C3}) into Eq.\ (\ref{eqA:dL-def1}),  we have 
\begin{equation}
  \left ( \frac{ H_{0} }{ c } \right )   d_{L}  
      =   (1+z)  \int_{1}^{1+z}  \frac{dy} {\left ( 1-  \frac{ C_{3} }{ C_{1} }  \right )   y^{ C_{1}}   +  \frac{ C_{3} }{ C_{1} }   } , 
\label{eqA:dLC1C3}  
\end{equation}
where $C_{1} > 0$ is assumed.
We can calculate the luminosity distance $d_{L}$ from Eq.\ (\ref{eqA:dLC1C3}). 
As a typical example \cite{Koma4}, the luminosity distance for $C_{3} =0$ is given by
\begin{equation}
  \left ( \frac{ H_{0} }{ c } \right )   d_{L}  =  \begin{cases}
         \frac{1+z} {C_{1} -1 } \left [ 1- (1+z)^{ -C_{1} +1 } \right ]   &   (C_{1} \neq 1) ,  \\ 
         (1+z)\ln (1+z)                                                                &   (C_{1} = 1)      .
\end{cases}
\label{eq:dLC1(C3=0)}  
\end{equation}

In this section, we obtain solutions for the combined model, assuming the single-fluid-dominated universe.
The solutions can be widely used because the modified Friedmann and acceleration equations simultaneously include both $H^{2}$ and $H$ terms.
In the next section, using the obtained solutions, 
we discuss two entropic-force models derived from the Bekenstein and generalized entropies.

In the present study, in order to examine entropic cosmology, we consider a spatially flat universe ($k = 0$), where $k$ is a curvature constant.
However, in a spatially non-flat universe ($k \neq 0$), the apparent horizon, $r_{A} = c/\sqrt{H^2 + (k /a^2) }$, does not coincide with the Hubble horizon, $r_{H}=c/H$, because of $k \neq 0$ \cite{Easson1}.
Accordingly, we use the apparent horizon as the preferred screen rather than the Hubble horizon, when we consider the spatially non-flat universe. 
Of course, in the standard entropic cosmology, it was argued that the extrinsic curvature at the surface was likely to result in an expression such as $\alpha_1 = \beta_1 = \frac{3}{2 \pi}$ and $\alpha_2 = \beta_2 = \frac{3}{4 \pi}$ \cite{Easson2,Koivisto1}.
Similarly, the influence of the curvature may be included in four coefficients $\alpha_1$, $\beta_1$, $\alpha_3$, and $\beta_3$.
Note that, in the present paper, we consider a spatially flat universe.
(Bulk viscous cosmology in the spatially non-flat universe has been examined in detail. See, e.g., Refs.\ \cite{Barrow12,Avelino2}.)

\section{Evolution of the universe in the two entropic-force models}
\label{Comparison}

As entropic-force terms, $H^{2}$ terms are derived from the Bekenstein entropy, whereas $H$ terms are derived from a generalized entropy. 
Thus, we propose two entropic-force models derived from the Bekenstein and generalized entropies.
We hereinafter refer to these models as the Bekenstein entropic-force model [Eqs.\ (\ref{eq:SmFRW01s_summ})--(\ref{eq:alpha_summ})] 
and the generalized entropic-force model [Eqs.\ (\ref{eq:mFRW01(nonadd)})--(\ref{eq:alpha3_beta3_s})], respectively.
In this section, we examine the evolution of the universe in the two entropic-force models.

\begin{table}[t]
\caption{Dimensionless constants for the two entropic-force models. 
The Bekenstein and Generalized columns indicate the information for the entropic-force models derived from the Bekenstein and generalized entropies, respectively. For details, see the text. }
\label{tab-parameter}
\newcommand{\m}{\hphantom{$-$}}
\newcommand{\cc}[1]{\multicolumn{1}{c}{#1}}
\renewcommand{\tabcolsep}{1.5pc} 
\renewcommand{\arraystretch}{1.25} 
\begin{tabular}{@{}lllll}
\hline
\hline
$\textrm{Parameter}$       &   $\textrm{Bekenstein}$         &   $\textrm{Generalized}$   \\
\hline
$\alpha_{1}$                      &  $0$                                 & $0$   \\
$\alpha_{3}$                      &  $0$                                 & $0$   \\
$\beta_{1}$                        &  $0.75(=3/4)$                   & $0$   \\
$\beta_{3}$                        &  $0$                                 & $0.884$   \\
\hline
$C_{1}$                              &  $0.75(=3/4)$                   & $1.5(=3/2)$   \\     
$C_{3}$                              &  $0$                                 & $0.884$   \\    
\hline
\hline
\end{tabular}\\
 \end{table}

For the two entropic-force models, we consider the matter-dominated universe given by
\begin{equation}
  w=0 . 
\label{eq:w=0}
\end{equation}
In Sec.\ \ref{combined model}, we obtained solutions for the combined model 
using four dimensionless constants, i.e., $\alpha_1$, $\alpha_3$, $\beta_1$, and $\beta_3$.
In other words, by setting the four constants, we can obtain the solution of each model.
The four constants for each model are listed in Table\ \ref{tab-parameter}.

For the Bekenstein entropic-force model, $\alpha_1$, $\alpha_3$, and $\beta_3$ are $0$, as described in Sec.\ \ref{Entropic-force model for the Bekenstein entropy} (and shown in Table \ref{tab-parameter}). 
Therefore, substituting these values and $w=0$ into Eq.\ (\ref{eq:C3}), we have $C_3 =0$.
On the other hand, $\beta_{1}$ is set to be $3/4$, in order to obtain $C_{1} =3/4$.
This is because, as reported in Ref.\ \cite{Koma4}, when we consider the equation $dH/dN = - C_{1} H$, the solution with $C_{1} =3/4$ was consistent with the observed supernova data.  
Here, $C_{1} =3/4$ was calculated from the Hawking temperature description \cite{Easson1}, without using a fitting with the supernova data \cite{Koma4}.
Of course, we accept that $\beta_{1}$ should be a free parameter.
For example, $C_{1}$ is approximated as $0.780$, if a fitting method discussed later herein is used to determine $C_{1}$. 
The value of $0.780$ is close to $3/4$, or, more specifically, to $\frac{3}{2}(1- \frac{3}{2 \pi}) = 0.7838 \dotsi $.
In Ref.\ \cite{Koma4}, $ C_{1} = \frac{3}{2}(1- \frac{3}{2 \pi})$ was calculated from the anticipated surface term order \cite{Easson1}.
We have confirmed that the properties of the universe for $ C_{1} = \frac{3}{2}(1- \frac{3}{2 \pi})$ or $ C_{1} = 0.780$ are similar to the properties for $ C_{1} = 3/4$.
Accordingly, in the present paper, we select $3/4$ as $C_{1}$ for the Bekenstein entropic-force model.

In contrast, for the generalized entropic-force model, $\alpha_1$, $\alpha_3$, and $\beta_1$ are $0$, as examined in Sec. \ref{Entropic-force model for a generalized entropy} (and shown in Table \ref{tab-parameter}). 
Substituting these values and $w=0$ into Eq.\ (\ref{eq:C1}), we obtain $C_{1} =3/2$.  
(In the standard cosmology, the universe for $C_{1} =3/2$ corresponds to the matter-dominated universe \cite{Weinberg1,Roy1,Ryden1}.)
On the other hand, $\beta_{3}$ is a free parameter. 
Therefore, $\beta_{3}$ is determined through fitting with a fine-tuned standard $\Lambda$CDM model based on the Planck 2013 results \cite{Planck2013}.
To this end, we use the luminosity distance. 

For the two entropic-force models, we calculate the luminosity distance $d_L$ from Eqs.\ (\ref{eqA:dLC1C3}) and (\ref{eq:dLC1(C3=0)}) and Table \ref{tab-parameter}.
For the standard $\Lambda$CDM model, the luminosity distance of the spatially flat universe is given as 
\begin{align}
 \left ( \frac{H_0}{c} \right )   d_{L} & =  (1+ z) \int_{0}^{z} dz' [ (1+z')^2 (1+ \Omega_{m} z')  \notag  \\
                                                   & \quad    -z'(2+z') \Omega_{\Lambda}]^{-1/2}   , 
\label{eq:dL(CDM)}
\end{align}
where $\Omega_{m} = \frac{\rho_{m}}{\rho_c} = \frac{8\pi G \rho_m}{3H_{0}^2}$ and $\Omega_{\Lambda}= \frac{\Lambda}{3 H_{0}^2}$ \cite{Carroll01}.
$\Omega_{m}$ and $\Omega_{\Lambda}$ represent the density parameters for matter and $\Lambda$, respectively. 
Moreover, $\rho_{c}$ represents the critical density, and $\rho_{m}$ is the density of matter, which includes baryon and dark matter.
The universe in which $(\Omega_{m}, \Omega_{\Lambda}) = (0.315, 0.685)$ is a fine-tuned standard $\Lambda$CDM model, which takes into account the recent Planck 2013 best fit values \cite{Planck2013}. 
Note that we assume $\Omega_{\textrm{total}}=  \Omega_{m} + \Omega_{\Lambda} = 1$ and neglect the density parameter $\Omega_{r}$ for the radiation \cite{Koma4,Koma4bc}.  

In the present study, $\beta_{3}$ for the generalized entropic-force model is determined through fitting with the fine-tuned standard $\Lambda$CDM model.
In fact, substituting $\alpha_{3}=0$ into Eq.\ (\ref{eq:C3}),  $C_{3}$ is equal to $\beta_{3}$. 
Accordingly, instead of $\beta_{3}$, we determine $C_{3}$ through fitting with the fine-tuned standard $\Lambda$CDM model, minimizing the following function \cite{Taru5} given by
\begin{equation}
\chi^{2} (C_{1}, C_{3}) = \sum\limits_{i=0}^{N_{z}} { \left[ \frac{   d_{L,\Lambda} (z)  - d_{L} (z; C_{1}, C_{3})  }{ d_{L,\Lambda} (z) }   \right]^{2}   }, 
\label{eq:min}
\end{equation}
where $d_{L,\Lambda} (z)$ and $d_{L} (z; C_{1}, C_{3})$ are the luminosity distances for the fine-tuned standard $\Lambda$CDM model and the combined (entropic-force) model, respectively.
For the generalized entropic-force model, $C_{1}$ is set to be $1.5$. 
In order to determine $C_{3}$, the redshift is varied from $z=0$ to $z=2$ in increments of $\Delta z = 0.01$.
In other words, $z$ is given by $ z = i \Delta z $, where $i$ is varied from $0$ to $N_{z}=200$.
Through fitting, $C_{3}$ is approximately determined to be $0.884$, as shown in Table \ref{tab-parameter}. 
Since $\beta_{3} = C_{3}$, $\beta_{3}$ is set to be $0.884$.
(It may be possible to consider $C_{1}$ and $C_{3}$ as free-parameters simultaneously. 
In this case, $C_{1}$ and $C_{3}$ are approximately $2.08$ and $1.58$, respectively.
We do not discuss such conditions in the present paper.)

\begin{figure} [t] 
\begin{minipage}{0.495\textwidth}
\begin{center}
\scalebox{0.3}{\includegraphics{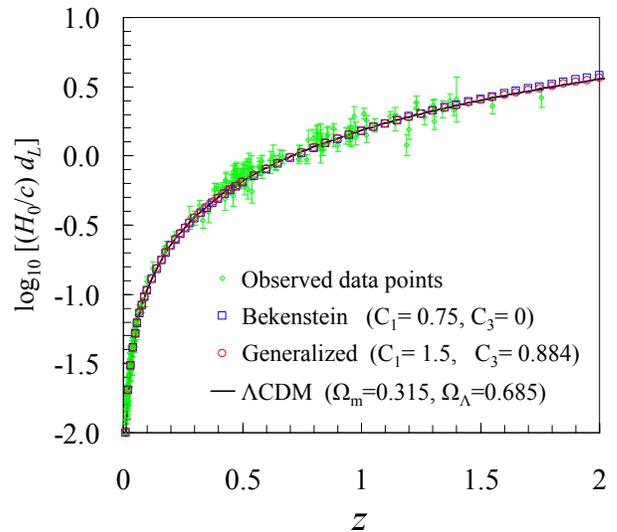}}
\end{center}
\end{minipage}
\caption{ (Color online). Dependence of luminosity distance $d_L$ on redshift $z$.
Here, Bekenstein and Generalized indicate the information for the entropic-force models derived from the Bekenstein and generalized entropies, respectively. The open diamonds with error bars are supernova data points taken from Ref.\ \cite{Riess2007SN1}. 
For supernova data points, $H_{0}$ is set to be $67.3$ km/s/Mpc  based on the Planck 2013 results \cite{Planck2013}. }
\label{Fig-dL-z}
\end{figure}

We now observe the luminosity distance $d_L$.
As shown in Fig.\ \ref{Fig-dL-z}, the two entropic-force models agree well with supernova data points and the fine-tuned standard $\Lambda$CDM model.
For the generalized entropic-force model, $C_{3}=0.884$ is determined through fitting with the fine-tuned standard $\Lambda$CDM model, as mentioned previously.
Therefore, it is not so surprising that the generalized entropic-force model agrees well with the standard $\Lambda$CDM model.
For the Bekenstein entropic-force model, $C_{1}= 0.75$ is selected, where the value of $0.75$ was calculated based on the description of the Hawking temperature \cite{Koma4}.  
However, it is demonstrated that the two entropic-force models can describe the present accelerating universe without adding the cosmological constant or dark energy.

As shown in Fig.\ \ref{Fig-dL-z}, the difference between the two entropic-force models is not clear when we observe the luminosity distance at late times. Therefore, we examine the time evolutions of the normalized scale factor  $a/a_{0}$ for longer time ranges.
For the two entropic-force models, we calculate $a/a_{0}$ from Eqs.\ (\ref{eq:a-t}) and (\ref{eq:a-t(C3=0)}) and Table \ref{tab-parameter}. 
For the standard $\Lambda$CDM model, we use the following equation \cite{Weinberg1,Ryden1}: 
\begin{equation}
 H_{0} (t - t_{0}) = \int_{1}^{\frac{a}{a_{0}}} { \frac{  d a^{\prime} }{   \sqrt {   \frac{ \Omega_{r} }{a^{\prime 2}}  + \frac{ \Omega_{m} }{ a^{\prime} } + \Omega _{\Lambda} a^{\prime 2}  + ( 1- \Omega_{\textrm{total}} ) }   }    }    .
\label{eq:a_CDM}
\end{equation}
We integrate Eq.\ (\ref{eq:a_CDM}) numerically in order to obtain the time evolution of the normalized scale factor $a/a_0$.
In the present study, for a spatially flat universe, we assume $\Omega_{\textrm{total}}=  \Omega_{r} + \Omega_{m} + \Omega_{\Lambda}  =1$ and neglect the influence of radiation, i.e., $\Omega_{r}=0$ \cite{Koma4}.

\begin{figure} [t]  
\begin{minipage}{0.495\textwidth}
\begin{center}
\scalebox{0.32}{\includegraphics{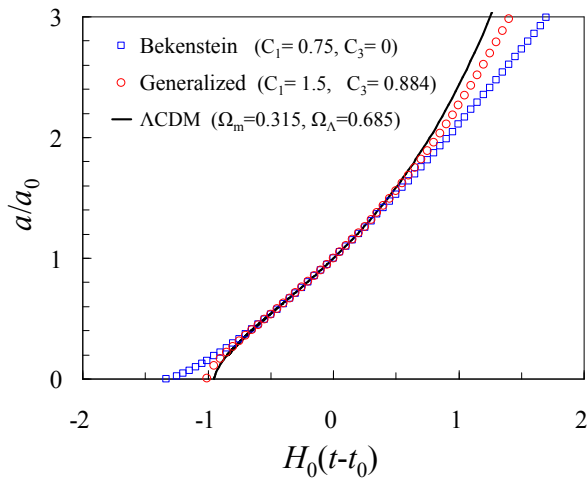}}
\end{center}
\end{minipage}
\caption{ (Color online). Time evolution of normalized scale factor $a/a_{0}$.
The horizontal axis is normalized as $H_{0} (t- t_{0})$. 
Here, Bekenstein and Generalized indicate the information for the entropic-force models derived from the Bekenstein and generalized entropies, respectively.}
\label{Fig-a-t}
\end{figure}

Figure \ref{Fig-a-t} shows the time evolutions of the normalized scale factor $a/a_0$.
For the Bekenstein entropic-force model, $a/a_0$ increases uniformly, because $a/a_0$ is given by $(C_{1} H_{0} t)^{1/C_{1}}$ [Eq. (\ref{eq:a-t(C3=0)})] when $C_{1} \neq 0$ and $C_{3}=0$. 
In fact, a temporal deceleration parameter $q(t)$ is given by $- \ddot{a}/(a H^{2}) = C_{1}  - (\hat{C}_{3}/H) -1$, as shown in Eq.\ (\ref{eq:ddota_a_C1C3h}).
Accordingly, $q(t)$ is constant, when $C_{1}$ is constant and $\hat{C}_{3} = C_{3} =0$.
In contrast, for the generalized entropic-force model, $q(t)$ can vary when $\hat{C}_{3} \neq 0$, even if $C_{1}$ is constant.
Therefore, we can expect that $a/a_0$ for the generalized model does not increase uniformly, unlike that for the Bekenstein model.
In order to observe this phenomenon, we focus on the generalized entropic-force model.
As shown in Fig.\ \ref{Fig-a-t}, the increase in $a/a_0$ tends to become gradually slower for $ H_{0}(t-t_{0}) \lessapprox -0.5$. However, for $ H_{0}(t-t_{0}) \gtrapprox -0.5$, the increase in $a/a_0$ tends to become gradually faster.
In other words, the generalized entropic-force model predicts a decelerated and accelerated expansion of the universe, which is similar to $a/a_0$ for the fine-tuned standard $\Lambda$CDM model.
Interestingly, the generalized entropic-force model can describe the decelerated and accelerated expansion, without adding a constant term such as a cosmological constant.

As discussed in Sec. \ref{Standard entropic-force models}, the standard entropic cosmology cannot describe a decelerating and accelerating universe \cite{Basilakos1,Sola_2013a}.
However, the standard entropic cosmology can describe the decelerating and accelerating universe 
if the modified Friedmann and acceleration equations include extra constant terms, as shown in Ref.\ \cite{Basilakos1}.
In the present study, we do not discuss such entropic-force models that include constant terms, because the origin of the constant terms is not clear.
(In $\Lambda(t)$CDM models, the constant terms are naturally obtained from an integral constant of the renormalization group equation for the vacuum energy density \cite{Sola_2013b}.)
Of course, bulk viscous cosmology can describe the decelerated and accelerated expansion \cite{Avelino2} because the modified acceleration equation includes the $H$ term as an extra driving term.
However, in the generalized entropic-force model, the $H$ term is derived from the generalized entropy, without using a bulk viscosity of cosmological fluids.

We now examine a transition time $t_{\textrm{trans}}$ \cite{Avelino2}, which represents the time of transition from the decelerated expansion epoch to the accelerated expansion epoch.
Calculating the second derivative of Eq.\ (\ref{eq:a-t}), equating to $0$, and solving the result, 
we obtain the transition time for the combined model as 
\begin{equation} 
H_{0} (t_{\textrm{trans}} -t_{0}) = \frac{1}{C_{3}} \ln (C_{1} - C_{3} )    \quad  \quad  (C_{1} > C_{3}>0)   .
\label{eq:trans-time}
\end{equation}
Substituting $C_{1}=1.5$ and $C_{3} =0.884$ into Eq.\ (\ref{eq:trans-time}), 
we have $H_{0} (t_{\textrm{trans}} -t_{0}) \approx -0.55$, for the generalized entropic-force model.
Note that $t_{\textrm{trans}}$ does not exist for the Bekenstein entropic-force model, 
because $a/a_{0}$ increases uniformly when $C_{3} =0$.

As shown in Fig.\ \ref{Fig-a-t}, the two entropic-force models are consistent with the fine-tuned standard $\Lambda$CDM model at the present time.
In order to discuss this more closely, we examine the deceleration parameter $q_{0}$.
For example, we can estimate $q_{0}$ for the fine-tuned standard $\Lambda$CDM model as $q_{0} \thickapprox  -0.53$, from $q_{0} = (\Omega_{m} -2 \Omega_{\Lambda} + 2 \Omega_{r} ) /2 $ \cite{Weinberg1}, 
where $(\Omega_{m}, \Omega_{\Lambda}, \Omega_{r})$ is set to be $(0.315, 0.685, 0)$.
For the two entropic-force models, $q_{0}$ is calculated from $q_{0} = C_{1} - C_{3} -1$, as shown in Eq.\ (\ref{eq:q0C1C3}).
Consequently, for the Bekenstein and generalized entropic-force models, we have $ q_{0} = -0.25$ and $ q_{0} \thickapprox   -0.38$, respectively.
Therefore, at the present time, the acceleration for the two entropic-force models is slightly slower than that for the fine-tuned standard $\Lambda$CDM model.

\begin{figure} [t]  
\begin{minipage}{0.495\textwidth}
\begin{center}
\scalebox{0.32}{\includegraphics{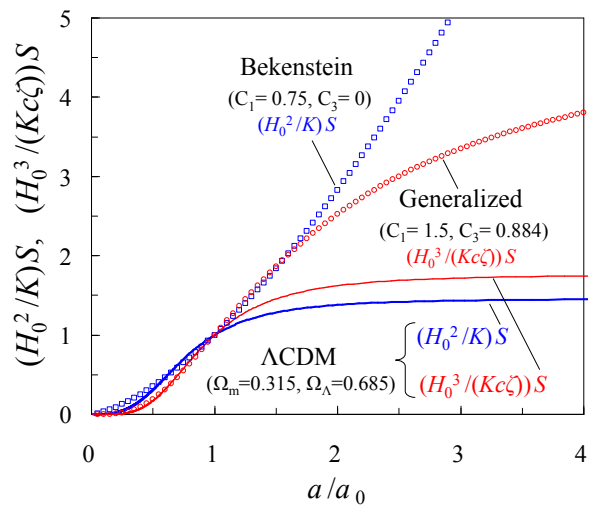}}
\end{center}
\end{minipage}
\caption{ (Color online). Evolutions of the Bekenstein and generalized entropies.
The vertical axis represents $ (H_{0}^{2} / K) S $ and $ (H_{0}^{3} / (K c \zeta) ) S $, for the Bekenstein and generalized entropic-force models, respectively.
The solid lines represent $(H_{0}^{2} / K) S$ and $(H_{0}^{3} / (K c \zeta) ) S $ for the fine-tuned standard $\Lambda$CDM model and are numerically calculated from $ (H/H_{0})^{-2}$ and $ (H/H_{0})^{-3} $.  }
\label{Fig-S-a}
\end{figure}

Finally, we observe the evolutions of the Bekenstein and generalized entropies.
In order to calculate the entropy, $S = K/ H^{2} $ [Eq.\ (\ref{eq:SH(add)2})] is used for the Bekenstein entropic-force model, 
whereas $S = Kc \zeta/ H^{3} $ [Eq. (\ref{eq:S(nonadd)2})] is used for the generalized entropic-force model.
In the following, the two entropies are normalized as $ (H_{0}^{2} / K) S $ and $ (H_{0}^{3} / (K c \zeta) ) S $, respectively.
We numerically calculate $ (H/H_{0})^{-2}$ and $ (H/H_{0})^{-3} $ of the fine-tuned standard $\Lambda$CDM model, 
in order to compare the two entropic-force models with the $\Lambda$CDM model.
This is because $ (H/H_{0})^{-2}$ and $ (H/H_{0})^{-3} $ correspond to $(H_{0}^{2} / K) S$ and $(H_{0}^{3} / (K c \zeta) ) S $, respectively.

Figure\ \ref{Fig-S-a} shows the evolutions of the Bekenstein and generalized entropies.
We first focus on the entropies for the fine-tuned standard $\Lambda$CDM model.
For $a/a_{0} \lessapprox 1$, the entropies for the standard $\Lambda$CDM model increase rapidly, 
whereas, for $a/a_{0} \gtrapprox  1$, the increase in the entropies tends to become gradually slower.
Since the standard $\Lambda$CDM model has been widely examined, similar results have been reported \cite{Easther1,Barrow3,Davies11_Davis0100,Gong00_01,Setare1,Cline1,Egan1}.
We now examine the two entropic-force models.
For $a/a_{0} \lessapprox  1$, the entropy for the Bekenstein entropic-force model is consistent with $(H_{0}^{2} / K) S$ for the standard $\Lambda$CDM model.
Similarly, for $a/a_{0} \lessapprox  1$, the entropy for the generalized entropic-force model agrees well with $(H_{0}^{3} / (K c \zeta) ) S $ for the $\Lambda$CDM model.
This indicates that the evolutions of the late universe for the two entropic-force models are consistent with the evolution for the $\Lambda$CDM model. 
We have confirmed this from Figs.\ \ref{Fig-dL-z} and \ref{Fig-a-t} as well.
However, for $a/a_{0} \gtrapprox  1$, the entropy for the Bekenstein entropic-force model increases uniformly, whereas the increase in the entropy for the generalized entropic-force model tends to become gradually slower [Fig.\ \ref{Fig-S-a}].
Therefore, for $a/a_{0} \gtrapprox 1$, the entropy for the Bekenstein entropic-force model is clearly different from 
$(H_{0}^{2} / K) S$ for the standard $\Lambda$CDM model. 
On the other hand, for $a/a_{0} \gtrapprox 1$, the entropy for the generalized entropic-force model increases more rapidly than $(H_{0}^{3} / (K c \zeta) ) S $ for the $\Lambda$CDM model.
However, the increase in the entropy for the generalized entropic-force model tends to become gradually slower. Accordingly, the evolution of the entropy for the generalized entropic-force model is similar to the evolution for the $\Lambda$CDM model.
This is because the generalized entropic-force model can describe a decelerating and accelerating universe.

\section{Discussion and conclusions}
\label{Conclusions}
In order to examine entropic cosmology, we have derived extra entropic-force terms not only from the Bekenstein entropy but also from a generalized entropy,  
assuming that the horizon of the universe has an entropy and a temperature. 
Because of nonadditive generalizations, the generalized entropy is proportional to volume, unlike for the Bekenstein entropy.
Consequently, as entropic-force terms, $H$ terms are derived from the generalized entropy, whereas $H^{2}$ terms are derived from the Bekenstein entropy.

Interestingly, the $H$ term is similar to an extra driving term for bulk viscous cosmology.
This similarity may be interpreted as a sign that the generalized entropy behaves as if it were a classical entropy generated by bulk viscous stresses.
This is probably because the generalized entropy considered here is proportional not to area but rather to volume. However, as one interpretation, the generalized entropy may be related to a bulk viscosity of cosmological fluids, through a holographic screen.
This interpretation does not contradict the following description; i.e., the bulk viscosity is considered to be the only thing that can generate an entropy in the homogeneous and isotropic universe.
In the present study, we assume an entropy on the horizon of the universe, using the holographic principle.
However, if we assume another entropy that is proportional to the volume of the universe, 
its entropic-force can likely explain the accelerated expansion of the universe.
In other words, the concept of the entropic-force may be applied to the bulk of the universe, without using the holographic principle.

In the present paper, based on effective pressure, we have formulated the modified Friedmann, acceleration, and continuity equations for two entropic-force models derived from the Bekenstein and generalized entropies.
The Friedmann equation is shown not to include the entropic-force term, whereas the continuity equation has a non-zero right-hand side. 
In order to examine the properties of the universe in the two entropic-force models, 
we have considered a combined model, in which the Friedmann and acceleration equations include both $H^{2}$ and $H$ terms simultaneously as the entropic-force terms.
We have obtained solutions of the combined model, assuming a homogeneous, isotropic, spatially flat universe, focusing on the single-fluid-dominated universe. 

We have confirmed that the two entropic-force models can describe the present accelerating universe, without adding the cosmological constant or dark energy. 
The Bekenstein entropic-force model is found to predict a uniformly accelerating universe, 
whereas the generalized entropic-force model predicts a decelerating and accelerating universe, as in the case for a fine-tuned standard $\Lambda$CDM model.
Similar properties of the universe with the extra driving terms, i.e., $H^{2}$ and $H$ terms, have been examined and discussed in not only bulk viscous cosmology but also $\Lambda(t)$CDM models with a variable cosmological term $\Lambda(t)$.
However, in the present study, the extra driving terms are derived from the Bekenstein and generalized entropies, without using a bulk viscosity of cosmological fluids and dark energy.

The present study has revealed fundamental properties of the expanding universe in entropic cosmologies based on the Bekenstein and generalized entropies.
In particular, the entropic-force derived from the generalized entropy will help in discussing the accelerating universe in other cosmological models.
For example, when an entropy is assumed to be proportional to $r_{H}^{4}$, the obtained entropic-force term is a constant, as if it were a cosmological constant.
Although such an entropy has probably not yet been suggested, we can discuss cosmological models from various entropic-force viewpoints.
In the present study, in order to examine entropic cosmology, we assume an entropy on the horizon of the universe.
At least in principle, it is possible to assume an entropy that is proportional to the volume of the universe in order to discuss entropic-force models.

In the present study, we have not examined cosmological fluctuations, since we have focused on background evolutions of the late universe.
However, through linear density perturbations, it has been discussed that bulk viscous models are difficult to reconcile with astronomical observations of structure formations \cite{Barrow21}.
In addition, as shown in Ref.\ \cite{Sola_2009}, $\Lambda(t)$CDM models similar to the combined model are not consistent with the structure formation data. 
(The structure formation in $\Lambda(t)$CDM models has been examined in detail \cite{Sola_2011a}.)
The previous works imply that the present entropic-force models are difficult to reconcile with the structure formation data. 
We expect that another type of entropy, e.g., $S \propto r_{H}^{4}$, helps to overcome the difficulty. 
This task is left for future research.


\begin{thebibliography}{99}
\bibitem{C_21}
We can obtain an alternative continuity equation from the first law of thermodynamics.
According to Ref.\ \cite{Koma4}, using Eq. (\ref{eq:S(nonadd)}), the modified continuity equation is calculated as
$ \dot{\rho} + 3 (\dot{a}/a) [ \rho   +  (p/c^2) ]  = -  \gamma  \frac{3}{4 \pi G} \frac{3 c \zeta}{2}  \dot{H} $.
This equation agrees with Eq. (\ref{eq:fluid_(nonadd)}), when $\dot{H} = - H^{2}$. 
In the present paper, we do not discuss the continuity equation obtained from the first law of thermodynamics.
\end{thebibliography}
\end{document}